\documentclass[12pt]{article}
\usepackage{amsmath,amssymb,epsfig}
\usepackage{graphicx,floatflt,subfigure}
\usepackage{cite}

\makeatletter

\renewcommand\section{\@startsection {section}{1}{\z@}%
                                   {-3.5ex \@plus -1ex \@minus -.2ex}%
                                   {2.3ex \@plus.2ex}%
                                   {\normalfont\large\bfseries}}

\renewcommand\subsection{\@startsection{subsection}{2}{\z@}%
                                     {-3.25ex\@plus -1ex \@minus -.2ex}%
                                     {1.5ex \@plus .2ex}%
                                     {\normalfont\normalsize\bfseries}}

\newcommand{\anti}[1]{\mbox{$\overline{\rm #1}$}}

\def\Im{\mathop{\mathrm{Im}}\nolimits}

\def\Re{\mathop{\mathrm{Re}}\nolimits}

\def\Tr{\mathop{\mathrm{Tr}}\nolimits}
\def\tr{\mathop{\mathrm{tr}}\nolimits}

\def\half{{\frac{1}{2}}}

\def\p{\partial}

\def\unit{{1\kern-.65ex {\rm l}}}
\def\1{{1\kern-.65ex {\rm l}}}

\def\vb{{\overline{v}}}

\def\zb{{\overline{z}}}

\def\taub{{\overline{\tau}}}

\def\CO{{\cal O}}

\def\bbC{{\mathbb{C}}}

\newcount\hour \newcount\minute
\hour=\time \divide \hour by 60
\minute=\time
\def\now{%
\ifnum \hour<13
  \ifnum \hour=0 \advance \hour by 12 \number\hour:\else \number\hour:\fi%
     \ifnum \minute<10 0\fi%
     \number\minute%
\ A.M.%
\else \advance \hour by -12 \number\hour:%
  \ifnum \minute<10 0\fi%
  \number\minute%
  \ P.M.%
\fi%
}

\makeatother

\begin{document}

\baselineskip=18pt  
\numberwithin{equation}{section}  



%
%


\thispagestyle{empty}

\vspace*{-2cm} 
\begin{flushright}
{\tt arXiv:0801.2154}\\
CALT-68-2668\\
ITFA-2008-01
\end{flushright}

\vspace*{0.8cm} 
\begin{center}
{\LARGE Off-shell M5 Brane, Perturbed Seiberg-Witten Theory, and Metastable Vacua\\}
 \vspace*{1.5cm}
 Joseph Marsano$^1$, Kyriakos Papadodimas$^2$, and Masaki Shigemori$^2$\\
 \vspace*{1.0cm} 
 $^1$ California Institute of Technology 452-48, Pasadena, CA 91125, USA\\[1ex]
 $^2$ 
 Institute for Theoretical Physics, University of Amsterdam\\
 Valckenierstraat 65, 1018 XE Amsterdam, The Netherlands\\
 \vspace*{0.6cm} 
 \verb|marsano_at_theory.caltech.edu|,
 \verb|kpapado_at_science.uva.nl|,
 \verb|mshigemo_at_science.uva.nl|
\end{center}
\vspace*{.5cm}

\noindent
We demonstrate that, in an appropriate limit, the off-shell M5-brane
worldvolume action effectively captures the scalar potential of
Seiberg-Witten theory perturbed by a small superpotential and,
consequently, any nonsupersymmetric vacua that it describes.
This happens in a similar manner to the emergence from M5's of the scalar potential describing certain type IIB flux configurations \cite{ourpaper}.
We then construct exact nonholomorphic M5 configurations in
the special case of $SU(2)$ Seiberg-Witten theory deformed by a
degree six superpotential which correspond to the recently discovered
metastable vacua of Ooguri, Ookouchi, Park \cite{Ooguri:2007iu}, and
Pastras \cite{Pastras:2007qr}.
These solutions take the approximate form of a holomorphic Seiberg-Witten
geometry with harmonic embedding along a transverse direction and allow
us to obtain geometric intuition for local stability of the gauge theory vacua.
As usual, dynamical processes in the gauge theory, such as the decay of 
nonsupersymmetric vacua, take on a different character in the M5 description
which, due to issues of boundary conditions, typically involves runaway behavior
in MQCD.

\newpage
\setcounter{page}{1} 



\tableofcontents


\section{Introduction}

String theory has a rich history of providing geometric intuition for
the structures that appear in supersymmetric field theories.  For
example, $NS5/D4$ constructions in type IIA and their corresponding
$M$-theory lifts naturally give rise to the Riemann surfaces
\cite{Witten:1997sc, Witten:1997ep, Giveon:1998sr} which play such a
crucial role on the gauge theory side \cite{Seiberg:1994rs,
Seiberg:1994aj}.  In this manner, the relation of geometry to the
structure and properties of supersymmetric vacua is made completely
manifest.

Since the recent discovery \cite{Intriligator:2006dd} that
supersymmetric gauge theories often admit metastable SUSY-breaking
vacua, significant effort has been devoted to the study of their stringy
realizations
\cite{Franco:2006es,GE:2006rw,Ooguri:2006pj,Ooguri:2006bg,Franco:2006ht,Bena:2006rg,Ahnstuff,Argurio:2006ny,Aganagic:2006ex,Tatar:2006dm,Kitano:2006xg,Heckman:2007wk,Giveon:2007fk,Argurio:2007qk,Murthy:2007qm,GE:2007vh,Kawano:2007ru,Douglas:2007tu,ourpaper,Malyshev:2007yb,Tatar:2007za,Aharony:2007pr,Heckman:2007ub,Aharony:2007db,Aganagic:2007kd,Kumar:2007dw,Mazzucato:2007ah,Aganagic:2007py,Giveon:2007ef,Giveon:2007ew,Buican:2007is,Tatar:2007wp}.
This is of particular interest for two
reasons.  First, one might hope to learn about the role played by
geometry in the structure of nonsupersymmetric vacua in gauge theory.
Second, stringy embeddings of metastable vacua can potentially provide
new means by which SUSY-breaking can be achieved in string theory.  Such
constructions are typically local in nature and are thus well-suited to
the sort of stringy model building advocated in \cite{Verlinde:2005jr}.

One potential pitfall is that classical brane constructions in type IIA
and their $M$-theory lifts can be well-studied only in parameter regimes
that are far from those in which the corresponding gauge theory
description is valid.  As such, there is no guarantee at the outset that
string theory knows anything about physics away from supersymmetric
vacua, where quantities are protected.  Despite this fact, considerable
progress has been made along this direction in the case of the ISS vacua
of ${\cal{N}}=1$ SQCD \cite{Intriligator:2006dd} and a number of
generalizations \cite{Ooguri:2006bg, Franco:2006ht, Bena:2006rg,
Tatar:2006dm, Giveon:2007fk, Argurio:2007qk, Tatar:2007za,
Giveon:2007ef, Giveon:2007ew,Tatar:2007wp}, and it has been demonstrated
that type IIA/M-theory realizations provide valuable intuition for the
metastability of those vacua.

In this paper, we shall focus instead on the stringy realization of
Seiberg-Witten theory deformed by a superpotential.  Because one can
tune the degree to which ${\cal{N}}=2$ SUSY is broken, the K\"ahler
potential is under some degree of control.  This not only allows one to
reliably compute the scalar potential in the gauge theory, it also gives
us hope that the standard stringy realizations might allow us to
engineer the full potential, in a suitable sense, along with any
non-SUSY vacua that it describes.

As is well-known by now, deformed Seiberg-Witten theory admits two
possible low energy descriptions depending on the various scales
involved in the problem.  If the characteristic scale, $g$, of the
superpotential that controls the mass of the adjoint scalar is
sufficiently small compared to the dynamical scale, $\Lambda$, of the
gauge group, then the superpotential only plays a role in the deep IR
after one has moved to the effective Abelian theory on the Coulomb
branch.  This is the situation referred to above in which the K\"ahler
potential is under tunably good control.  On the other hand, if $g$ is
much larger than $\Lambda$, we should integrate out the adjoint scalar
before passing to the IR\@.  The ${\cal{N}}=2$ SUSY is broken to
${\cal{N}}=1$ and the gauge group is Higgs'ed at a high scale, below
which non-Abelian factors confine.
Certain aspects of the low energy dynamics, such as chiral condensate
and the value of superpotential in supersymmetric vacua, are captured by
an effective theory for the corresponding glueball superfields
$S_i\sim\frac{1}{32\pi^2}\text{tr}\,W_{\alpha\,i}^2$
\cite{Veneziano:1982ah, Cachazo:2002ry} and corrections from the massive
adjoint scalar lead to generation of the Dijkgraaf-Vafa superpotential
\cite{Dijkgraaf:2002fc, Dijkgraaf:2002vw, Dijkgraaf:2002dh,
Dijkgraaf:2002xd}.

As pointed out by \cite{Vafa:2000wi} and \cite{Aganagic:2006ex} in the context of
the type IIB realization of this story via large $N$ duality
\cite{Cachazo:2001jy, Cachazo:2001gh, Cachazo:2001sg}, the appearance of
FI terms suggests that the glueball effective theory actually possesses
a spontaneously broken ${\cal{N}}=2$ supersymmetry.  This again gives
one control over the K\"ahler potential and allows a scalar potential to
be reliably computed{\footnote{The result of \cite{ourpaper} that the
resulting potential also arises in a type IIA description provides
evidence for the structure of spontaneously broken ${\cal{N}}=2$ SUSY in
the case of IIB on local Calabi-Yau in the presence of flux.  
}}.  Quite remarkably, the IIA brane construction which
realizes deformed Seiberg-Witten theory admits an $M$-theory lift that
incorporates and unifies the scalar potentials of both regimes.  In
particular, each can be obtained from a different limit of the $M5$
worldvolume action and hence both follow from simple minimal area
equations.  That the Dijkgraaf-Vafa potential is accurately encoded was
essentially derived in \cite{ourpaper}.  In this paper, we will
explicitly demonstrate how the Seiberg-Witten potential can also be
realized in a suitable limit {\footnote{That it should be possible to
realize this potential is not surprising because it has been
well-established that the M5 realization of Seiberg-Witten theory
correctly captures the K\"ahler potential and fails to reproduce
nonholomorphic quantities only at four derivative order and higher
\cite{deBoer:1997zy}.}}.

As an application of this result, we then proceed to describe stringy
embeddings of the metastable vacua of deformed Seiberg-Witten theory
recently discovered by Ooguri, Ookouchi and Park \cite{Ooguri:2007iu}
and Pastras \cite{Pastras:2007qr}, which we will refer to as OOPP vacua.
More specifically, we will find exact minimal area M5's which reduce in
the appropriate limit to minima of the Seiberg-Witten potential.  By
taking a IIA limit of these configurations, we will also be able to
obtain a simple geometric picture for these vacua analogous to that of
\cite{Ooguri:2006bg, Franco:2006ht, Bena:2006rg} for the ISS vacua
\cite{Intriligator:2006dd}.  This will allow us to obtain a
semiclassical understanding from the $NS5/D4$ point of view of the full
scalar potential itself, as well as the stability of the
supersymmetry-breaking vacua.  To our knowledge, the resulting mechanism
of SUSY-breaking in brane constructions is novel.

While this work was in progress, a different approach to the realization
of OOPP vacua appeared \cite{Mazzucato:2007ah} which incorporates the
superpotential of \cite{Ooguri:2007iu, Pastras:2007qr} in a slightly
different manner.  While it can be written in a single trace form, the
superpotential of \cite{Ooguri:2007iu, Pastras:2007qr} nonetheless has
degree larger than the rank of the gauge group and hence can also be
viewed as a multitrace object.  The authors of \cite{Mazzucato:2007ah}
took this point of view and attempted to realize the OOPP theory by
using extra NS5-branes to directly engineer a multitrace superpotential.
We take a different approach motivated by the $T$-dual type IIB
constructions where a single-trace superpotential determines the
geometry and the rank of the gauge group simply specifies the number of
$D5$ branes that can occupy singular points.  From this point of view,
the background geometry remains the same regardless of how many or how
few $D5$ branes we wish to add.  As such, we realize the OOPP
superpotentials by simply curving the NS5-branes in the same manner that
we would for an identical superpotential with gauge group of arbitrarily
high rank.  That we are able to obtain the Seiberg-Witten potential from
the M5 worldvolume action in this way seems to justify the use of this
approach for engineering aspects of the OOPP theory away from the
supersymmetric vacua.  It would be interesting to study whether this
might also be possible in the setup of \cite{Mazzucato:2007ah}{\footnote{As we shall see, the manner in which the scalar potential arises in our setup suggests that one must relax some assumptions of \cite{Mazzucato:2007ah} in order to see it in that context.}}

The organization of this paper is as follows.
We begin in section \ref{sec:IIASW} with a brief review of the type
IIA/M construction of deformed Seiberg-Witten theory.
In section \ref{sec:DV}, we then review the mechanism described in
\cite{ourpaper} by which this construction is able to realize not only
the space of supersymmetric vacua, but also the full scalar potential
for the light degrees of freedom in the Dijkgraaf-Vafa regime.
We then perform a similar analysis in the Seiberg-Witten regime in
section \ref{sec:SW} to demonstrate that the IIA/M picture captures the
scalar potential there as well.
In section \ref{sec:OOPP}, we review the manner in which
nonsupersymmetric vacua can be engineered in the Seiberg-Witten regime
by choosing a suitable superpotential \cite{Ooguri:2007iu,
Pastras:2007qr}.
In section \ref{sec:curves}, we specialize to the case of $SU(2)$ and construct exact minimal-area $M5$ configurations corresponding to these vacua.   
In section \ref{sec:mechanism}, we study their semiclassical limit and obtain a simple 
interpretation in the NS5/D4 language of the potential and mechanism by which stability
is achieved.
We close with some concluding remarks in section \ref{sec:conclusion}.
The Appendices include various technical details that we use in the main
text.
%

\section{Type IIA/M Description of Deformed Seiberg-Witten Theory}\label{sec:IIASW}

\begin{figure}
\begin{center}
\subfigure[NS5/D4 realization of the classical Seiberg-Witten moduli space.]
{\epsfig{file=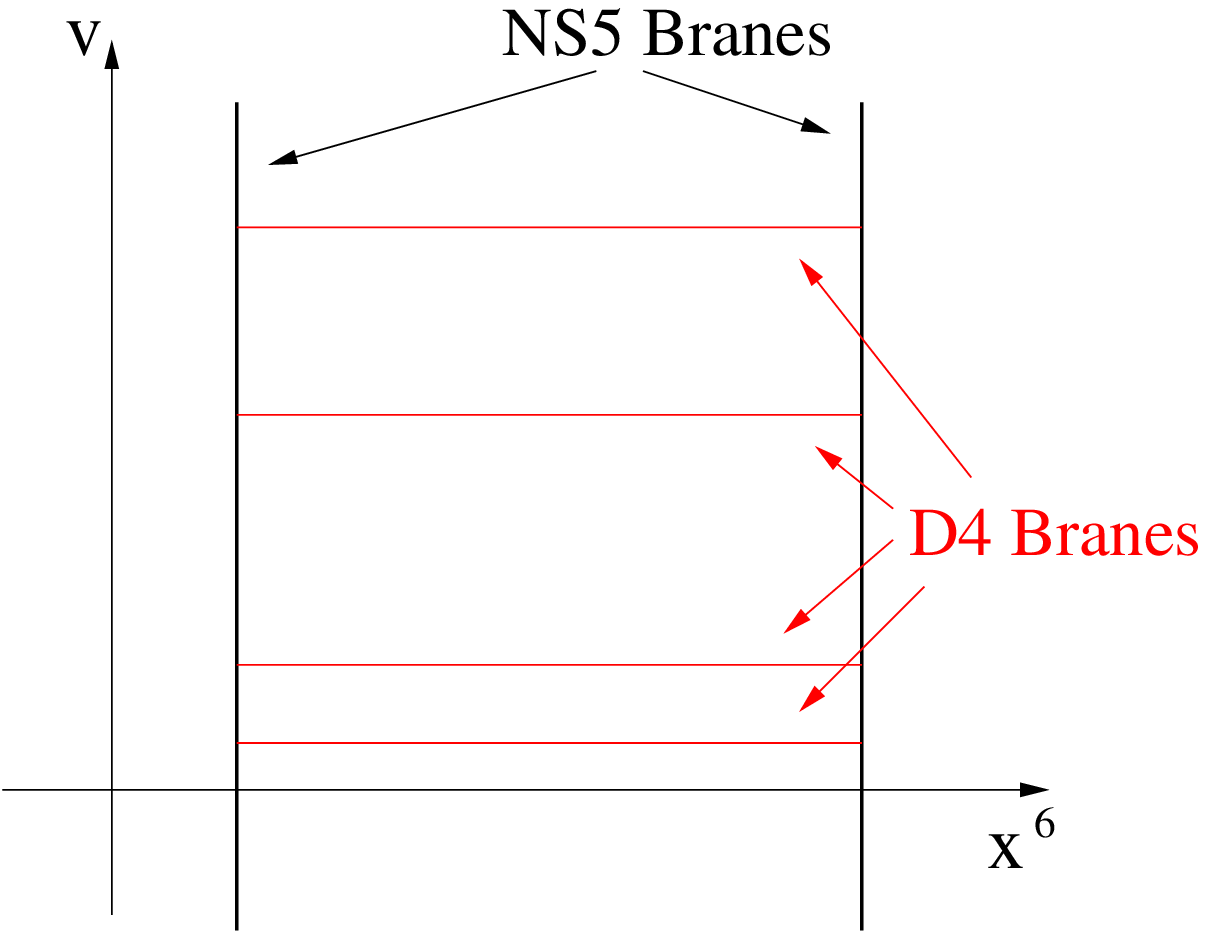,width=0.45\textwidth}\label{SWNS5D4}}
\hspace{0.1\textwidth}
\subfigure[Cartoon of M5 lift which realizes the quantum Seiberg-Witten moduli space.]
{\epsfig{file=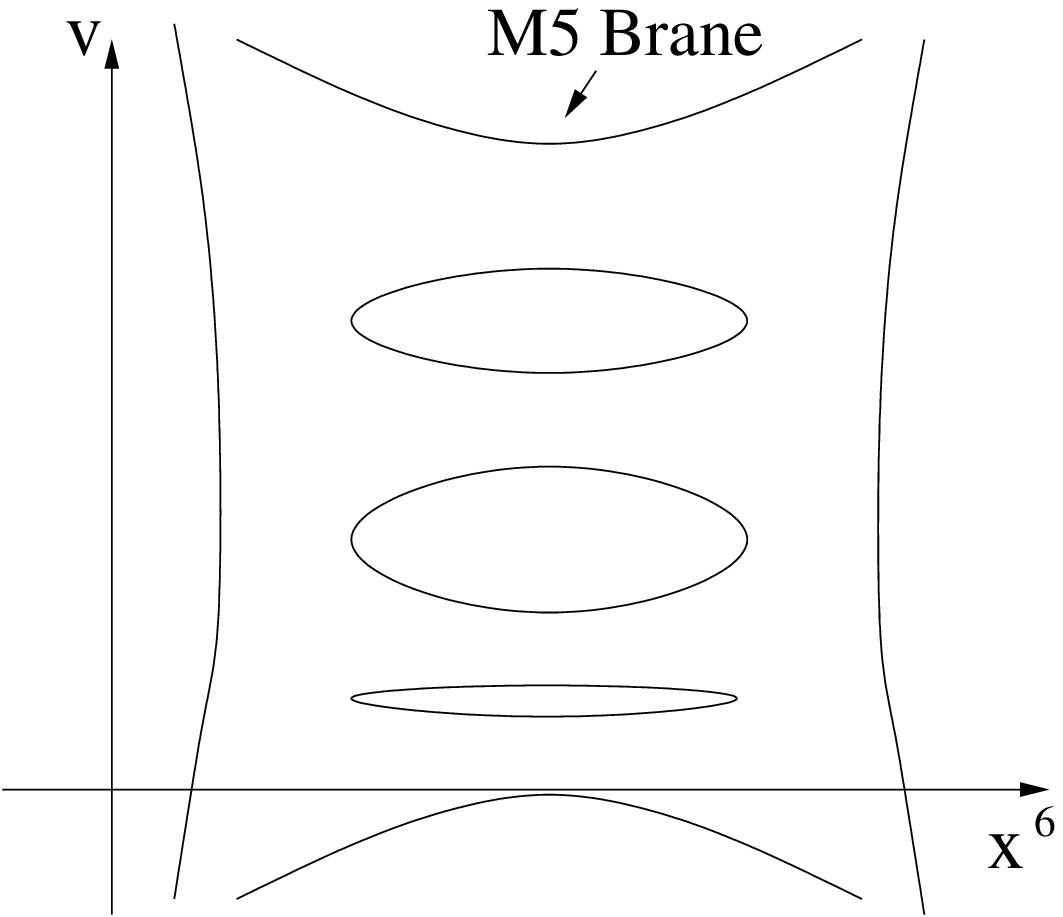,width=0.4\textwidth}\label{SWM5}}
\label{IIASW}\caption{Type IIA/M realization of classical and quantum moduli space of Seiberg-Witten theory.}
\end{center}
\end{figure}

Let us begin with a brief review of the type IIA/M realization of $SU(N)$ ${\cal{N}}=2$ supersymmetric Yang-Mills theory deformed by a superpotential for the adjoint scalar.  
As first shown by Witten \cite{Witten:1997sc}, the theory without superpotential
can be engineered in type IIA by starting with two NS5-branes and $N$ D4-branes extended along the 0123 directions.  The NS5's are further
extended along a holomorphic direction parametrized by the combination
\begin{equation}v=x^4+ix^5\end{equation}
and the D4's are suspended between them along $x^6$ as depicted in
figure \ref{SWNS5D4}.  If we scale the NS5 separation $L$ along $x^6$ to
zero, the D4 worldvolume becomes effectively four-dimensional with gauge
coupling constant given by
\begin{equation}\frac{8\pi^2}{g_{YM}^2}=\frac{L}{g_s\sqrt{\alpha'}}.\end{equation}
The resulting configuration preserves 8 supercharges, so this
worldvolume theory is nothing other than the
celebrated ${\cal{N}}=2$ $SU(N)$ supersymmetric Yang-Mills theory studied
by Seiberg and Witten \cite{Seiberg:1994rs}.

\subsection{Classical and Quantum Moduli Space from Geometry}

Because of the protection afforded by supersymmetry, we expect that the
moduli space of vacua can be seen directly in the brane constructions,
even away from the strict $L\rightarrow 0$ limit.  In figure
\ref{SWNS5D4}, for example, the classical moduli space of Seiberg-Witten
theory is evident in our ability to place the D4's at arbitrary values
of $v$.  To describe the quantum moduli space, however, it is necessary
to accurately treat the NS5/D4 intersection points.  While the strength
of the string coupling in the NS5 throat makes this difficult to do
directly in type IIA, Witten pointed out \cite{Witten:1997sc} that one
can make progress by noting that NS5's and D4's are two different
manifestations of the same object, namely the M5 brane of M-theory.  As
a result, the configuration of figure \ref{SWNS5D4} should be replaced
by a single M5 with four directions along 0123 and the remaining two, in
the probe approximation, extended on a nontrivial two-dimensional
surface $\Sigma$ of minimal area.  From the IIA point of view, this will
take the form of a curved NS5-brane with flux{\footnote{For the
discussion of supersymmetric vacua, we can be a bit careless about
validity of the probe approximation but a detailed description of the
necessary conditions, which are needed when considering
nonsupersymmetric configurations, can be found in \cite{ourpaper}.}}.

\begin{floatingfigure}{0.5\textwidth}
\begin{center}
\vspace{0.5cm}
\epsfig{file=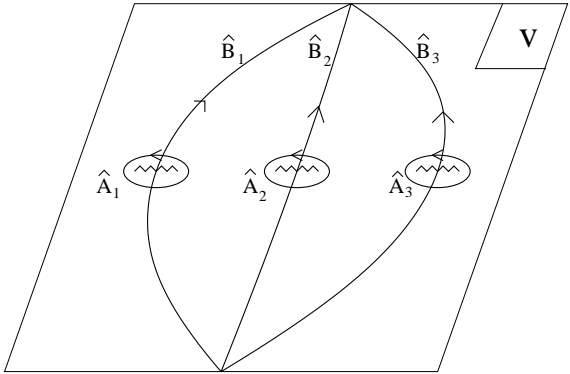,width=0.45\textwidth}
\caption{Sample parametrization of the M5 curve as double-cover of $v$-plane for the case $n=3$ with $\hat{A}$ and $\hat{B}$ cycles indicated.}
\vspace{2cm}
\label{vplane}
\end{center}
\end{floatingfigure}

The general structure of the M5 curve associated to figure \ref{SWNS5D4}
can be seen by observing that, roughly speaking, the stacks of D4-branes
blow up into tubes as illustrated in figure \ref{SWM5}.  Consequently, a
point on the classical moduli space where the D4's form $n$ distinct
stacks will correspond to a Riemann surface $\Sigma$ of genus $n-1$ with
two punctures{\footnote{The punctures represent the points at infinity
from which RR flux associated to the D4's can flow.}}.  A convenient
parametrization of this surface is as a double-cover of the $v$-plane
with $n$ cuts{\footnote{This relies on the fact that $\Sigma$ is
hyperelliptic \cite{Witten:1997sc, Witten:1997ep}.}}, as depicted in
figure \ref{vplane}.  Roughly speaking, we can think of each sheet as an
NS5 and the cuts as the tubes associated to the D4's.  In this case, the
full embedding is specified by providing the $v$-dependence of $x^6$ and
$x^{10}$, which are naturally paired into the complex combination
\cite{Witten:1997sc}
\begin{equation}s=R_{10}^{-1}\left(x^6+ix^{10}\right),\end{equation}
where 
\begin{equation}R_{10}=g_s\sqrt{\alpha'}\end{equation} 
is the radius of the M-theory circle.  We will set $\alpha'=1$ henceforth.

The embedding relevant for figure \ref{SWNS5D4} must be holomorphic due
to the supersymmetry and is determined by imposing boundary conditions
appropriate for the system at hand.  In particular, we must fix the
separation of the NS5's at some cutoff scale $\Lambda_0${\footnote{This
is essentially equivalent to fixing the bare coupling constant in the
Yang-Mills theory at a UV cutoff scale.  It should be noted that the
brane configurations under study are UV completed into MQCD and not
simply an asymptotically free gauge theory so it will only make sense to
compare with the gauge theory at scales below this cutoff.}} and specify
the number of D4's in each stack.  These conditions can be conveniently
summarized in terms of constraints on the periods of the 1-form $ds$
around the $\hat{A}$ and $\hat{B}$ cycles of figure \ref{vplane}
{\footnote{Note that there are $n$ $\hat{A}$-cycles and $n$
$\hat{B}$-cycles despite the fact that the curve has genus $n-1$.  This
allows us to treat the curve with marked points as a degenerate Riemann
surface of genus $n$, putting meromorphic 1-forms with single poles,
such as $ds$, on an equal footing with holomorphic 1-forms.}}
\begin{equation}\frac{1}{2\pi i}\oint_{\hat{A}_j}ds = N_j\qquad\text{and}\qquad \frac{1}{2\pi i}\oint_{\hat{B}_j}ds = -\alpha_j.\label{SWbdryconds}\end{equation}
Here $N_j$ is the number of D4-branes in the $j$th stack and
\begin{equation}\alpha_j = -\frac{4\pi i}{g_{YM}^2}+\frac{\theta_j}{2\pi}\label{alphadef}\end{equation}
with $g_{YM}^2$ denoting the bare coupling at scale $\Lambda_0$.  In
general, one can consider nontrivial relative $\theta_j$ angles by
allowing the $\alpha_j$ to differ by integers.  From this point onward,
however, we shall restrict for simplicity to the case in which all
$\alpha_j$ are equivalent:
\begin{equation}\alpha_j\equiv \alpha\quad\text{for all }j.\end{equation}

To find the family of curves which satisfy the constraints
\eqref{SWbdryconds} for various choices of $N_j$, Witten
\cite{Witten:1997sc} noted that since the periods of $ds$ take integer
values, the coordinate
\begin{equation}t\equiv \Lambda^{N} e^{-s}\label{tdef}\end{equation}
must be well-defined on the curve.  With this observation, he
demonstrated that the M5 lifts are described by nothing other than the
Seiberg-Witten geometries
\begin{equation}t^2-2P_N(v)t+\Lambda^{2N}=0,\label{SWcurves}\end{equation}
where $P_N(v)$ is a polynomial of degree $N$ and
\begin{equation}\Lambda^{2N}=\Lambda_0^{2N}e^{2\pi i\alpha}\end{equation}
is the dynamical scale of the gauge group.  Generic choices of $P_N(v)$
lead to (degenerate) genus $N$ surfaces and hence correspond to lifts of
classical configurations in which all $D4$-branes are separated.  For
special choices of $P_N(v)$, however, the curves \eqref{SWcurves}
degenerate into surfaces of lower genus which describe the lifts of
configurations with some $N_j> 1$.

\subsection{Turning on a Superpotential}

\begin{figure}
\begin{center}


\epsfig{file=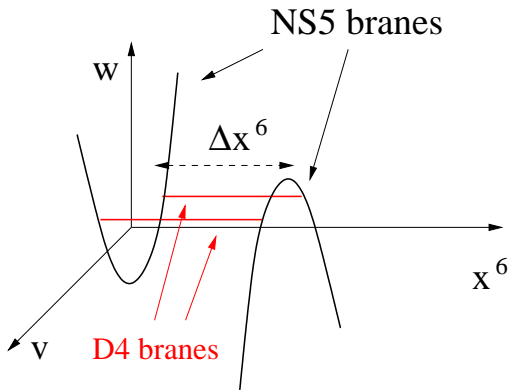,width=0.4\textwidth}
\caption{Sample NS5/D4 realization of supersymmetric vacua in $U(N)$ ${\cal{N}}=2$ supersymmetric Yang-Mills with a cubic superpotential.}

\label{A1cubTdualD4}
\end{center}
\end{figure}

To similarly engineer the ${\cal{N}}=2$ theory deformed by a polynomial
superpotential $W_n(\Phi)$ of degree $n+1$, we need only modify the
NS5/D4 construction of figure \ref{SWNS5D4} by curving the NS5-branes
appropriately \cite{Witten:1997ep, Hori:1997ab, deBoer:1997ap}.
More specifically, we introduce the complex combination
\begin{equation}w=x^7+ix^8\end{equation}
and extend the NS5's along the holomorphic curves $w(v)= \pm W_n'(v)$.
An example of such a configuration for the case of a cubic
superpotential is depicted in figure \ref{A1cubTdualD4}.  Classically,
the supersymmetric vacua correspond to configurations where the
D4-branes sit at zeros of $W_n'(v)$, in accordance with our expectations
from the gauge theory side.


At quantum level, the supersymmetric vacua are again effectively captured by moving to the M5 lift.  The only new ingredient here is the addition of an extra nontrivial embedding coordinate whose $v$-dependence we must specify.  The boundary conditions appropriate for that coordinate are determined by the asymptotic geometry of the NS5's, meaning that we must impose
\begin{equation}w(v)\sim \pm W_n'(v)\end{equation}
near the points at infinity.  Holomorphic solutions to this constraint take the form
\begin{equation}w(v)=\sqrt{W_n'(v)^2-f_{n-1}(v)}\label{wvcurves}\end{equation}
for $f_{n-1}(v)$ a generic polynomial of degree $n-1$.  If we further require consistency of the $w(v)$ and $s(v)$ embeddings, we are led to the factorization formulae
\begin{equation}P_N(v)^2-4\Lambda^{2N}=S_{N-n}(v)^2\left(W_n'(v)^2-f_{n-1}(v)\right)\label{facteqn}\end{equation}
if $n<N$ and similarly
\begin{equation}\left(P_N(v)^2-4\Lambda^{2N}\right)H_{n-N}(v)^2=W_n'(v)^2-f_{n-1}(v)\label{facteqnn}\end{equation}
if $n>N$.
For given $W_n'(v)$, there are at most a finite number of choices for
$P_N(v)$, $f_{n-1}(v)$, and $S_{N-n}(v)$ ($H_{n-N}(v)$), which satisfy
\eqref{facteqn} (\eqref{facteqnn}).  This reflects the familiar fact
that adding a superpotential to Seiberg-Witten theory lifts all but a
discrete set of points on the moduli space.  That these conditions yield
vacua in agreement with the gauge theory analysis has been
well-established \cite{Cachazo:2002pr, Cachazo:2002zk} and will also
fall out naturally from our general formalism to follow.

\section{The Dijkgraaf-Vafa Regime}\label{sec:DV}

We now turn to a study of physics away from the supersymmetric vacua.
As discussed in the introduction, the appropriate IR description depends
strongly on the relative sizes of the characteristic mass $g$ of the
adjoint scalar field and the dynamical scale $\Lambda$.  In this
section, we consider the case where $g$ is sufficiently large compared
to $\Lambda$.

\subsection{Review of the Gauge Theory Side}

On the gauge theory side, when $g$ is sufficiently large the gauge group
is Higgs'ed and ${\cal{N}}=2$ supersymmetry broken to ${\cal{N}}=1$ at a
high scale.  The remaining non-Abelian factors confine and certain
aspects of the IR dynamics can be captured by the glueball superfields
$S^i=-\frac{1}{32\pi^2}\tr W_{\alpha}^{i\,2}$ as well as the
${\cal{N}}=1$ gauge multiplets corresponding to the overall $U(1)$'s
{\footnote{Note that the situation is quite subtle when Abelian factors
remain after Higgs'ing because the stringy constructions still contain
glueballs for them \cite{Intriligator:2003xs}.  We will largely avoid
this technicality here and refer the interested reader to
\cite{Intriligator:2003xs}, where this issue is discussed in greater
detail.}}.  The leading contribution to the glueball superpotential is
given by the Veneziano-Yankielowicz term \cite{Veneziano:1982ah} and
corrections arising from the presence of the adjoint scalar can be
computed by integrating it out order by order in perturbation theory
\cite{Dijkgraaf:2002xd}.  This calculation receives contributions only
from planar diagrams which can in turn be computed with an auxiliary
holomorphic matrix model
\begin{equation}Z=\int\,{\cal{D}}\Phi\,\exp\left(-\frac{1}{g_s}\tr W_n(\Phi)\right).\label{matmod}\end{equation}
In the end, one arrives at the famous Dijkgraaf-Vafa superpotential
\cite{Dijkgraaf:2002fc, Dijkgraaf:2002vw, Dijkgraaf:2002dh}
\begin{equation}W = \alpha_iS^i+N^i\frac{\partial {\cal{F}}}{\partial S^i}(S^j),\label{DVsup}\end{equation}
where $N^i$ denote ranks of the confining non-Abelian factors, $\alpha_i$ are as in \eqref{alphadef} and ${\cal{F}}$ is the planar contribution to the free energy of \eqref{matmod} from the saddle point where $N^i$ eigenvalues of $\Phi$ sit at the $i$th critical point of $W_n(\Phi)$.
Note that in writing ${\cal{F}}$ as a function of $S^i$, we must make the identification $S^i\sim g_sN^i$. This Dijkgraaf-Vafa superpotential  was also derived from anomaly arguments in \cite{Cachazo:2002ry}.

As usual, the matrix model computation can be reinterpreted in a
geometric language based on the corresponding spectral curve.  In the
case at hand, this auxiliary Riemann surface takes the form
\begin{equation}w^2=W_n'(v)^2-f_{n-1}(v)\label{speccurv}\end{equation}
and the function ${\cal{F}}$ is simply its prepotential.  To make this
more precise, let us use the double-cover of the $v$-plane to
parametrize \eqref{speccurv} and specify $\hat{A}$ and $\hat{B}$ cycles
as in figure \ref{vplane}.  The quantities $S^i$ and
$\partial{\cal{F}}/\partial S^j$ are then given by suitable $\hat{A}$
and $\hat{B}$ period integrals
\begin{equation}S^j=\frac{1}{2\pi i}\oint_{\hat{A}_j}\,w\,dv\qquad\qquad \frac{\partial{\cal{F}}}{\partial S^j}=\frac{1}{2\pi i}\oint_{\hat{B}_j}\,w\,dv.\end{equation}
As usual, $\partial_j\partial_k{\cal{F}}\equiv \hat{\tau}_{jk}\,${\footnote{We use the notation $\hat{\tau}_{ij}$ instead of $\tau_{ij}$ for the period matrix of a generic hyperelliptic curve.  The reason for this is to avoid confusion later then $\tau$ is used as the complex structure modulus for an auxiliary torus.}} yields the period matrix of the Riemann surface \eqref{speccurv}.

These geometric formulae arise quite naturally when this theory is
engineered in type IIB using $D5$-branes wrapping singular 2-cycles of a
local Calabi-Yau \cite{Vafa:2000wi, Cachazo:2001jy, Cachazo:2001gh,
Cachazo:2001sg}.  There, large $N$ duality relates the brane setup to a
deformed Calabi-Yau with flux, whose superpotential takes the precise
form \eqref{DVsup}.  In this context, it has been conjectured that the
IR physics contains an underlying ${\cal{N}}=2$ structure
\cite{Vafa:2000wi, Aganagic:2006ex}, which gives one control over the
K\"ahler metric.  Such a possibility is suggested by the observation
that the IR degrees of freedom naturally combine into $n$ ${\cal{N}}=2$
vector multiplets and the superpotential \eqref{DVsup} is simply a
linear combination of electric and magnetic Fayet-Iliopoulos parameters.
If the conjecture is true, it would imply that we have some control over
the K\"ahler metric and, in particular, that it can be identified with
the imaginary part of the period matrix, $\Im\hat{\tau}_{ij}$, of
\eqref{speccurv}.  In that case, the scalar potential takes a relatively
simple form
\begin{equation}V_{DV}=\overline{\left(\alpha_i+N^k\hat{\tau}_{ki}\right)}\left(\Im\hat{\tau}\right)^{ij}\left(\alpha_j+\hat{\tau}_{j\ell}N^{\ell}\right).\label{DVpot}\end{equation}

In the next subsection, we will review the result of \cite{ourpaper}
that this potential is naturally encoded in the type IIA/M realization
of the gauge theory.  The resulting configuration is actually related to
the IIB setup described above by a $T$-duality \cite{Dasgupta:2001um,
Oh:2001bf}.  Because the IIA/M and IIB descriptions are only reliable in
widely-separated regions of parameter space, agreement of the scalar
potentials was far from guaranteed and provides significant indirect
evidence that, at least in the string picture, some residual
${\cal{N}}=2$ supersymmetry may be present.


\subsection{Type IIA/M Description} \label{subsec:DVM}

From the point of view of our NS5/D4 constructions, the Dijkgraaf-Vafa
regime has the natural interpretation as one where the D4's are
essentially pinned to the zeroes of $W_n'(v)$.  This means that the low
energy modes involve not motion of the D4's, but rather fluctuations in
the size of the tubes into which they blow up in the M5 lift.  In other
words, this is a regime where the $w(v)$ part of the curve is
essentially rigid while the $s(v)$ embedding can fluctuate.

Such a limit is in fact quite natural because the $s$ coordinate comes
with a natural scale, namely the radius of the M-circle
$R_{10}=g_s\sqrt{\alpha'}$.  If we take $R_{10}$ to be small, in which case
our minimal area M5 instead has the interpretation of a curved NS5-brane
with flux, we expect that variations of the $s$ coordinate of the
embedding actually comprise the lightest excitations of the system.  To
make this more precise, let us introduce a ``worldsheet'' coordinate,
$z$, to parametrize the nontrivial part of the M5 and define our
embedding by the functions
\begin{equation}s(z,\bar{z}),\quad v(z,\bar{z}),\quad w(z,\bar{z}).\end{equation}
In the probe approximation, the worldvolume theory is described by the Nambu-Goto action
\begin{equation}S\sim \frac{1}{R_{10}^2}\int_{\Sigma}\,d^2z\,\sqrt{g(s,v,w)},\label{M5actDV}\end{equation}
where $g(s,v,w)$ is the induced metric on the ``worldsheet'' $\Sigma$.
If we suppose that $R_{10} ds \ll dw,dv$ then this action can be expanded to
quadratic order as
\begin{equation}
S\sim \frac{1}{R_{10}^2}\int_{\Sigma}\,d^2z\,\sqrt{\tilde{g}(v,w)}+\int_{\Sigma}\,|ds|^2+\dots,\label{M5actDVexp}
\end{equation}
where $\tilde{g}(v,w)$ is the induced metric that arises from the $v$
and $w$ parts of the embedding alone.  At small $R_{10}ds$, the dominant term
of this action is the first one, whose equations of motion constrain the
$w$ and $v$ coordinates to describe a minimal area embedding.  If we
further impose the holomorphic boundary conditions $w(v)\sim \pm
W_n'(v)$ near infinity, we are again led to the family of holomorphic
curves
\begin{equation}w^2=W_n'(v)^2-f_{n-1}(v).\label{wvDV}\end{equation}
Note that the $\sqrt{\tilde{g}}$ term of \eqref{M5actDVexp} has flat
directions corresponding to the complex moduli of \eqref{wvDV}.

Turning now to the second term of \eqref{M5actDV}, the equations of
motion for $ds$ imply that it must be a harmonic 1-form on \eqref{wvDV}.
Because we fix the $\hat{A}$ and $\hat{B}$ periods of $ds$ according to
\eqref{SWbdryconds}, though, this leads to a unique $ds$ for each curve
of the family \eqref{wvDV}.  In particular, the 1-form $ds$ that we
obtain exhibits explicit dependence on the complex moduli of
\eqref{wvDV} and hence, plugging this result back into the action, we
find that $\int_{\Sigma}\,|ds|^2$ serves as a potential on the space of
complex structures.  It is precisely this object that will correspond to
the scalar potential \eqref{DVpot}.

\subsubsection{Regularization of the Action}\label{subsubsec:regularization}

Our M5's are noncompact, though, so computations of their area must be
carefully regulated.  While we can do this quite easily by introducing a
cutoff $\Lambda_0$ along $v$, it is important to make sure that this
procedure leads to a result that is meaningful.  The issue of computing
regulated areas in this context has been discussed previously by de Boer
{\it et al.}\ \cite{deBoer:1998by}.  Our situation is essentially the
same as theirs because, in the regime under consideration, the
off-diagonal components of the induced metric $\tilde{g}(v,w)$ are
negligible.  Let us proceed to review their result, focusing for now
only on the evaluation of $\int_{\Sigma}^{\Lambda_0}\,|ds|^2$.

Because $ds$ is harmonic, it can be written as the sum of a holomorphic 1-form $ds_H$ and an antiholomorphic 1-form $d\bar{s}_A$
\begin{equation}ds = ds_H + d\bar{s}_A.\end{equation}
In this language, the object we want to compute is
\begin{equation}\begin{split}\int_{\Sigma}^{\Lambda_0}\,|ds|^2&=\frac{1}{2i}\int_{\Sigma}^{\Lambda_0}\,\left[ds_H\wedge d\bar{s}_H + ds_A\wedge d\bar{s}_A\right]\\
&=\frac{1}{2i}\int_{\Sigma}^{\Lambda_0}\,\left[\left(ds_H\wedge d\bar{s}_H-ds_A\wedge d\bar{s}_A\right)+2ds_A\wedge d\bar{s}_A\right]\\
&=\frac{1}{2i}\int_{\Sigma}^{\Lambda_0}\,\left[ds\wedge d\bar{s} + 2ds_A\wedge d\bar{s}_A\right].
\end{split}\end{equation}
We have chosen to write it in this manner because $ds\wedge d\bar{s}$ is
the restriction to $\Sigma$ of a closed 2-form in the target space.  For
this reason, its integral over $\Sigma$ should in fact be independent of
the moduli.\footnote{Another way to see that this integral must be taken
constant is to $T$-dualize the system into type IIB, where this integral
$\int ds\wedge d\bar{s}$ gets related to the integral of 3-form flux
$\int_{CY} G\wedge \bar{G}$, $G=F_{RR}+(i/g_s^{\rm
IIB})H_{NS}$\cite{ourpaper}.  Because this is a topological quantity, we
must keep this constant as we vary the moduli.} In practice, however,
the regulated integral of $ds\wedge d\bar{s}$ will depend on both the
moduli and the cutoff, $\Lambda_0$.  Consequently, to obtain a
meaningful regularization, we must choose the cutoff $\Lambda_0$ to vary
with the moduli in such a manner that the regulated quantity
\begin{equation}
\frac{1}{2i}\int_{\Sigma}^{\Lambda_0}\,ds\wedge d\bar{s}\label{regconst}
\end{equation}
is indeed constant on the moduli space.

With such a scheme in place, the potential $\int_{\Sigma}^{\Lambda_0}\,|ds|^2$ is simply given by
\begin{equation}\int_{\Sigma}^{\Lambda_0}\,|ds|^2 \sim \frac{1}{i}\int_{\Sigma}^{\Lambda_0}\,ds_A\wedge d\bar{s}_A,\label{regsarea}\end{equation}
where we have simply dropped the constant term \eqref{regconst}.  In the
work of de Boer {\it et al.}, the quantity analogous to \eqref{regsarea}
was cutoff-independent so this was the end of the story.  In our case,
however, this result may still exhibit a nontrivial dependence on the
cutoff.
Nevertheless, \eqref{regsarea} provides a suitable notion of regularized
area provided we also include the necessary moduli-dependence of
$\Lambda_0$.

\subsubsection{Evaluation of the Regulated $\int^{\Lambda_0}_{\Sigma}\,|ds|^2$}

Let us turn now to evaluation of \eqref{regsarea}.  We begin by using
the constraint \eqref{SWbdryconds} to obtain an expression for $ds$ in
terms of the period matrix $\hat{\tau}_{ij}$ of \eqref{wvDV}.  We write
$ds_H$ and $ds_A$ as
\begin{equation}ds_H = h^id\hat{\omega}_i\qquad\text{and}\qquad ds_A = \ell^id\hat{\omega}_i,\label{sexp}\end{equation}
where the $d\hat{\omega}_i$ comprise a basis of $n$ holomorphic
1-forms{\footnote{More precisely, the $d\hat{\omega}_i$ are meromorphic
1-forms with poles of degree at most 1 at $\infty$ on the two sheets.
They correspond to holomorphic 1-forms if we consider $\Sigma$ to be a
degenerate Riemann surface of genus $n$.}}, satisfying
\begin{equation}\frac{1}{2\pi i}\oint_{\hat{A}_j}\, d\hat{\omega}_i = \delta_{ij}.\label{canonnorm}\end{equation}
The $\hat{B}$-periods of $d\hat{\omega}_i$ yield elements of the period
matrix
\begin{equation}\frac{1}{2\pi i}\oint_{\hat{B}_j}\, d\hat{\omega}_i = \hat{\tau}_{ij}\end{equation}
and the constraints \eqref{SWbdryconds} become
\begin{equation}\begin{split}h^i-\bar{\ell}^i&=N^i,\\
h^i\hat{\tau}_{ij}-\bar{\ell}^i\bar{\hat{\tau}}_{ij}&=-\alpha_j.
\end{split}\end{equation}
This implies that
\begin{equation}\begin{split}
h^i&=-\frac{1}{2i}\left(\Im\hat{\tau}^{-1}\right)^{ij}\left(\alpha_j+\bar{\hat{\tau}}_{jk}N^k\right),\\
\bar{\ell}^i&=-\frac{1}{2i}\left(\Im\hat{\tau}^{-1}\right)^{ij}\left(\alpha_j+\hat{\tau}_{jk}N^k\right).
\end{split}\end{equation}
Now that we have determined $ds$, it is easy to evaluate
\begin{equation}\begin{split}\frac{1}{i}\int_{\Sigma}\,ds_A\wedge d\bar{s}_A &= -2\Im\left(\sum_j\oint_{\hat{A}_j}ds_A\,\oint_{\hat{B}_j}d\bar{s}_A\right)\\
&=8\pi^2\ell^i\left(\Im\hat{\tau}\right)_{ij}\bar{\ell}^j\\
&= 2\pi^2\overline{\left(\alpha_i+\hat{\tau}_{ik}N^k\right)}\left(\Im\hat{\tau}^{-1}\right)^{ij}\left(\alpha_j+\hat{\tau}_{j\ell}N^{\ell}\right)\\
&\sim V_{DV}.
\end{split}\end{equation}
This is precisely the scalar potential \eqref{DVpot}, as promised.

One potential pitfall in this calculation, though, is the fact that the
period matrix $\hat{\tau}_{ij}$ involves the computation of noncompact
$\hat{B}$ periods that must be regulated by $\Lambda_0$.  This means
that $\hat{\tau}_{ij}$ will in general depend on $\Lambda_0$ and hence
could exhibit additional dependence on the moduli in our regularization
scheme.  To see that this doesn't happen, note that
\begin{equation}\begin{split}\frac{1}{i}\int_{\Sigma}\,ds_H\wedge d\bar{s}_H &= -2\Im\left(\sum_j\oint_{\hat{A}_j}\,ds_H\,\oint_{\hat{B}_j}\,d\bar{s}_H\right)\\
&=8\pi^2h^i\left(\Im\hat{\tau}\right)_{ij}\bar{h}^j\\
&=2\pi^2\overline{\left(\alpha_i+\bar{\hat{\tau}}_{ik}N^k\right)}\left(\Im\hat{\tau}^{-1}\right)^{ij}\left(\alpha_j+\bar{\hat{\tau}}_{j\ell}N^{\ell}\right)\\
&= 2\pi^2\overline{\left(\alpha_i+\hat{\tau}_{ik}N^k\right)}\left(\Im\hat{\tau}^{-1}\right)^{ij}\left(\alpha_j+\hat{\tau}_{j\ell}N^{\ell}\right)-8N^j\left(\Im\alpha\right)_j.
\end{split}\end{equation}
The regulated integral of $ds\wedge d\bar{s}$ is therefore independent of the moduli
\begin{equation}2\int_{\Sigma}\,ds\wedge d\bar{s} = -8N^j\left(\Im\alpha\right)_j,\end{equation}
meaning that we can take our cutoff $\Lambda_0$ to be a large, moduli-independent constant.

\section{The Seiberg-Witten Regime}\label{sec:SW}

We now turn to the main subject of interest in this paper, namely the
Seiberg-Witten regime, where $g$ is sufficiently small that the superpotential can be
treated as a deformation of the IR effective description of
Seiberg and Witten \cite{Seiberg:1994rs}.  In what follows, we will
proceed to review some aspects of the gauge theory side.  We will then
turn to the type IIA/M description and demonstrate that the scalar
potential arises in a manner quite analogous to what we saw in the
Dijkgraaf-Vafa regime above.  This will further suggest that, by analogy
to \cite{ourpaper}, critical points in the Seiberg-Witten regime can be
associated to full M5 solutions in a suitable sense that we shall describe.

\subsection{Review of the Gauge Theory Side}

Before studying the superpotential deformation, let us first review some
aspects of the Seiberg-Witten effective description
\cite{Seiberg:1994rs} of the low energy physics.  The classical moduli
space of the theory is parametrized by eigenvalues $a^i$ of the adjoint
scalar, $\Phi$.  The quantum moduli space, on the other hand, is
equivalent to the moduli space of hyperelliptic curves of the form
\begin{equation}t^2-2P_N(v)t+\Lambda^{2N}=0,
\label{SWcurve}\end{equation} where $\Lambda$ is the dynamical scale of
the gauge group and
\begin{equation}P_N(v)=v^N-\sum_{k=2}^Ns_kv^{N-k}=\prod_{i=1}^N(v-a^i).\label{PNdef}\end{equation}
Here, the $s_k$ are symmetric polynomials of the $a^i$ and comprise one convenient parametrization of the moduli space.

\begin{figure}
\begin{center}
\epsfig{file=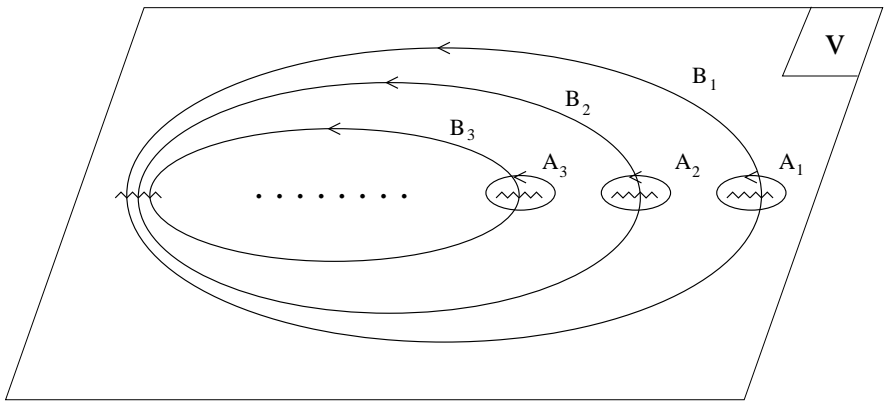,width=0.7\textwidth}
\caption{Depiction of Seiberg-Witten geometry \eqref{SWcurve} along with the basis of $A$ and $B$ cycles that we use to analyze the Seiberg-Witten regime..}
\label{SWvplane}
\end{center}
\end{figure}

As usual, we view \eqref{SWcurve} as a double-cover of the $v$ plane
with $n\le N$ branch cuts.  It is conventional to choose the cuts so
that they encircle pairs of branch points which coalesce in the
classical limit $\Lambda\rightarrow 0$.  As for 1-cycles in this
geometry, it will be useful in our analysis of the Seiberg-Witten regime
to use the basis of $A$ and $B$ cycles depicted in figure \ref{SWvplane}
as opposed to the $\hat{A}$ and $\hat{B}$ cycles introduced
previously{\footnote{This choice of basis differs from that used in our
analysis of the Dijkgraaf-Vafa regime by the removal of one $A$-cycle and the replacement of noncompact $\hat{B}$ cycles with compact ones.  This basis facilitates comparison to the traditional analysis of Seiberg-Witten theory and is also more naturally adapted to the boundary conditions we will use in section \ref{subsec:SWNS5D4}.}}
Note in particular that we have no need to introduce an $N$th $A$-cycle
because we will consider only 1-forms with vanishing residues in what
follows.

While the $s_k$ nicely describe the moduli space of curves \eqref{SWcurve}, a more convenient parametrization is given by expectation values of powers of $\Phi$
\begin{equation}u_p=\frac{1}{p}\left\langle\tr\Phi^p\right\rangle,\label{updef}\end{equation}
where $2\le p\le N$.  These are usefully encoded in the resolvent, a meromorphic 1-form with first order poles on the curve \eqref{SWcurve} defined as
\begin{equation}T(v)\,dv = \left\langle\tr\frac{dv}{v-\Phi}\right\rangle.\end{equation}
An explicit expression for this 1-form can be obtained by  noting that its $A$-periods and residue at $\infty$ are completely determined by the classical limit $\Lambda\rightarrow 0$.  In particular,
\begin{equation}\frac{1}{2\pi i}\oint_{A_j}T(v)\,dv = N_j,\qquad \frac{1}{2\pi i}\oint_{v=\infty}T(v)\,dv=N,\end{equation}
where $N_j$ is the multiplicity of the $j$th eigenvalue of $\Phi$ in the
classical limit.  Because a meromorphic 1-form on \eqref{SWcurve} with a
first order pole at $\infty$ is uniquely fixed by its $A$-periods and
residue, we arrive at the result\cite{deBoer:2004he}
\begin{equation}T(v)\,dv = -\frac{dt}{t}=ds,\end{equation}
where $s$ and $t$ are related as in \eqref{tdef}.  From this, one can easily compute $u_p$ via
\begin{equation}u_p=\frac{1}{2\pi ip}\oint_{v=\infty}v^p\,ds. 
\label{upcomp}\end{equation}

As for the eigenvalues $a^i$ themselves, they combine with their magnetic duals $a_{D\,j}$ to form a holomorphic section of an $Sp(2n,\mathbb{Z})$ bundle over the moduli space.  This section can be computed explicitly by integrating the Seiberg-Witten 1-form
\begin{equation}d\lambda_{SW}=v\,ds\end{equation}
about $A$ and $B$ cycles of the curve \eqref{SWcurve}  
\begin{equation}a^i=\oint_{A_i}\,d\lambda_{SW},\qquad
a_{D,j}=\oint_{B_j}d\lambda_{SW}.\end{equation}

In the absence of a superpotential, the low energy dynamics are
determined solely by the moduli space metric, which determines the
kinetic terms for the $u_p$
\begin{equation}
g^{r \bar{s}} = \frac{\partial a^i}{\partial u_r}\left(\Im\hat{\tau}\right)_{ij}\frac{\partial\bar{a}^j}{\partial\bar{u}_{\bar{s}}}.
 \label{kahlermet}
\end{equation}
We now return, however, to the situation at hand in which we deform the
theory by a nontrivial superpotential of the form
\begin{equation}W_n(\Phi)=\sum_{m=0}^{n}\frac{g_{m}\tr\Phi^{m+1}}{m+1}.\end{equation}
If the $g_m$ are all 
suitably small, we can treat this as a perturbation of the effective
theory whose precise form at a point $u_p$ of the moduli space is simply
given by its expectation value
\begin{equation}W_{\text{eff}}(u_p)=\left.\left\langle W_n(v)\right\rangle\right|_{u_p}=\sum_{m=0}^n \oint_{v=\infty} \frac{g_mv^{m+1}}{m+1}\,T(v)\,dv.\end{equation}
This superpotential, combined with the K\"ahler metric
\eqref{kahlermet}, completely determines the IR dynamics of the
deformed Seiberg-Witten theory in this regime.  Because we know both,
the scalar potential that controls the full vacuum structure of the
theory can be written down reliably as
\begin{equation}
V_{SW}=
\overline{\partial_{u_r}W_{\text{eff}}(u_p)}\overline{\left(\frac{\partial u_r}{\partial a^i}\right)}\left(\Im\hat{\tau}^{-1}\right)^{ij}\left(\frac{\partial u_s}{\partial a^j}\right)\partial_{u_s}W_{\text{eff}}(u_q),
\label{SWpot}
\end{equation}
which we refer to as the Seiberg-Witten potential.  It is this object
that we shall see arising from the type IIA/M description in the next
subsection.
 
Before moving on, though, let us note that while we have considered the $SU(N)$ situation above and will continue to focus on this example throughout the rest of this paper, it is easy to generalize to $U(N)$ by allowing $s_1\ne 0$, leading to an additional term of degree $v^{N-1}$ in \eqref{PNdef}.  All of the formalism readily generalizes.
 
\subsection{Type IIA/M Description}\label{subsec:SWNS5D4}
 
From the point of view of our NS5/D4 constructions, the Seiberg-Witten
regime has the natural interpretation as one where the curvature of the
NS5's is sufficiently small that the D4's can move along the $v$
direction away from zeroes of $W_n'(v)$ with very little cost in energy.
This suggests that configurations relevant for the Seiberg-Witten regime
are those whose lightest excitations correspond to changes in the $w$
embedding coordinate.  As such, they can be approximately described by
expanding the action \eqref{M5actDV} for small $dw$.  By analogy to
\eqref{M5actDVexp}, we find
\begin{equation}S\sim \frac{1}{R_{10}^2}\left(\int_{\Sigma}^{\Lambda_0}\,d^2z\,\sqrt{\tilde{g}(v,s)}+\int_{\Sigma}^{\Lambda_0}\,|dw|^2+\ldots\right),
\label{M5actSWexp}
\end{equation}
where $\tilde{g}(v,s)$ is the induced metric that arises from the $v$ and $s$ parts of the embedding alone.  When $dw$ is suitably small{\footnote{The precise condition is easy to work out in explicit examples, as we shall see later.}}, the dominant term of this action is the first one, whose equations of motion constrain the $v$ and $s$ coordinates to describe a minimal area embedding
Further imposing the conditions \eqref{SWbdryconds}, we again arrive at the Seiberg-Witten geometry \eqref{SWcurves}
 \begin{equation}t^2-2P_N(v)t+\Lambda^{2N}=0.
 \label{SWcurvess}\end{equation}
The complex structure moduli of \eqref{SWcurvess} correspond to flat directions of the first term in \eqref{M5actSWexp}.

The equations of motion for $dw$ which follow from the second term of \eqref{M5actSWexp} imply that it must describe a harmonic 1-form on the curve \eqref{SWcurvess}.  As we shall see in a moment, the boundary conditions for $w$ lead to a unique such 1-form for each curve of the family \eqref{SWcurvess}.  Plugging this back into \eqref{M5actSWexp}, we find that $\int_{\Sigma}^{\Lambda_0}\,|dw|^2$ serves as a potential on the space of complex structures.  It is precisely this object that will correspond to the scalar potential \eqref{SWpot}.

We now turn our attention to the computation of $\int_{\Sigma}^{\Lambda_0}\,|dw|^2$.  First, however, we must address the problem of finding the appropriate $dw$ for a given point in moduli space.  The boundary conditions that we impose are twofold.  First, we require as usual that
\begin{equation}w(v)\sim \pm W_n'(v)=\sum_{m=0}^ng_{m}v^m\label{SWwvbdry}\end{equation}
near $v=\infty$ on both sheets.  Second, however, is the condition that $w$ be single-valued.  This means that the integral of $dw$ over any compact period must vanish:
\begin{equation}\oint_{A_i}\,dw = 0\text{ for all }i,\qquad\oint_{B_j}dw = 0\text{ for all }j.\label{SWwper}\end{equation}

The next step is to expand $dw$ in a manner analogous to \eqref{sexp}.
Because of the polynomial behavior \eqref{SWwvbdry} at $\infty$, $dw$ is
comprised not only of holomorphic 1-forms but also meromorphic ones with
poles of degree 2 through $n+1$.{\footnote{Note that the absence of
logarithmic behavior at $\infty$ implies that $dw$ must have vanishing
residue at $\infty$.  It is for this reason that we did not introduce an
$N$th $A$-cycle in figure \ref{SWvplane} in order to treat the
Seiberg-Witten geometry as a degenerate Riemann surface of genus $N$.}}

Before moving on, let us make our choice of basis 1-forms a little more
precise.  To start, we consider the holomorphic 1-forms on
\eqref{SWcurvess}.  If we let $r_k$ denote some generic set of moduli,
which could be $s_k$, $u_k$, or some other suitable choice, a standard
collection of 1-forms is given by
\begin{equation}\frac{\left(\partial P_N(v)/\partial r_k\right)\,dv}{\sqrt{P_N(v)^2-\Lambda^{2N}}}=\frac{\partial}{\partial r_k}\left(s\,dv\right).\end{equation}
For example, if we use the $s_k$ in this construction, the result is the standard basis of 1-forms
\begin{equation}\frac{\partial}{\partial s_k}\left(s\,dv\right)=-\frac{v^{N-k}}{\sqrt{P_N(v)^2-\Lambda^{2N}}}.\end{equation}
For our purposes, we shall be more interested in the basis constructed from the $u_k$
\begin{equation}d\eta^k=\frac{\partial}{\partial u_k}\left(s\,dv\right).\end{equation}
The $d\eta^k$ are not canonically normalized but this is easily fixed.
In particular, if we define
\begin{equation}\sigma^{jk}\equiv\oint_{A_j}d\eta^k\end{equation}
then a collection of 1-forms $d\omega_i$ satisfying
\begin{equation}\frac{1}{2\pi i}\oint_{A_j}d\omega_i=\delta_{ij}\end{equation}
can be easily written as 
\begin{equation}d\omega_i = \sigma_{ik}^{-1}d\eta^k.\end{equation}
Note that we already see one piece of \eqref{SWpot} appearing
because{\footnote{Note that the boundary term that arises upon
integrating by parts gives no contribution because it is simply the
period of $ds$, which is a modulus-independent constant.}}
\begin{equation}
\sigma^{jk}=\frac{\partial}{\partial u_k}\oint_{A_j}s\,dv=-\frac{\partial}{\partial u_k}\oint_{A_j}v\,ds = -\frac{\partial}{\partial u_k}\oint_{A_j}\,d\lambda_{SW}=-\frac{\partial a^j}{\partial u_k}.
\label{sigmadef}\end{equation}

Now that we have discussed our explicit basis for holomorphic 1-forms,
we now turn to the meromorphic 1-forms of degrees 2 to $n+1$ that can
enter into $dw$.  These are often referred to as meromorphic
differentials of the second kind and in our situation generically take
the form
\begin{equation}\frac{Q(v)\,dv}{\sqrt{P_N(v)^2-\Lambda^{2N}}},\end{equation}
where $Q(v)$ is a polynomial of degree $\ge N$.  A convenient choice of
basis elements $d\Omega_m$, $m\ge 1$ is one for which the $A$-periods are
vanishing
\begin{equation}\oint_{A_i}d\Omega_m=0\label{OmegaAper}\end{equation}
and behave near $v=\infty$ as
\begin{equation}d\Omega_m = \left(\pm mv^{m-1}+{\cal{O}}\left(v^{-2}\right)\right)\,dv.\label{Omegaasymp}\end{equation}
To see that such a collection can indeed be constructed, let us start from generic meromorphic 1-forms $d\tilde{\Omega}_m$ of the form
\begin{equation}d\tilde{\Omega}_m = \frac{Q_{N+m-1}(v)\,dv}{\sqrt{P_N(v)^2-\Lambda^{2N}}}.\end{equation}
for $Q_{N+m-1}(v)$ a polynomial of degree $N+m-1$.  The constraint \eqref{OmegaAper} yields $N-1$ constraints while fixing the coefficients of all poles at $\infty$ yields another $m+1$.  In the end, this gives $N+m$ constraints, which is equivalent to the number of coefficients in the polynomial $Q_{N+m-1}(v)$.  Fortunately, we will not need to know the explicit form of $Q_{N+m-1}(v)$ in what follows, as only the leading behavior \eqref{Omegaasymp} is relevant for us.
 
We can now finally expand $dw$ as
\begin{equation}dw=dw_H+d\bar{w}_A\label{dw}\end{equation}
with
\begin{equation}dw_H = T_md\Omega_m +h^id\omega_i\qquad dw_A = R_p d\Omega_p+\ell^id\omega_i.\label{dwH_dwA}\end{equation}
The boundary condition \eqref{SWwvbdry} implies that
\begin{equation}T_m=g_m,\qquad\qquad \bar{R}_{p}=0,\end{equation}
while the vanishing of $A$ and $B$ periods \eqref{SWwper} leads to
\begin{equation}h^i=\bar{\ell}^i=\frac{1}{4\pi}\left(\Im\tau^{-1}\right)^{ij}K_{jm}g_m,\end{equation}
where
\begin{equation}K_{jm}=\oint_{B_j}d\Omega_m.\end{equation}
Our boundary conditions have thus led to a unique choice of $dw$ for each point on the moduli space.  Now turning to the evaluation of $\int_{\Sigma}^{\Lambda_0}\,|dw|^2$, we recall from our discussion of regularization in section \ref{subsubsec:regularization} that it is sufficient to evaluate
\begin{equation}\int_{\Sigma}^{\Lambda_0}\,|dw|^2\sim \frac{1}{i}\int_{\Sigma}^{\Lambda_0}\,dw_A\wedge d\bar{w}_A.\end{equation}
This is a straightforward task and leads to
\begin{equation}\begin{split}\frac{1}{i}\int_{\Sigma}^{\Lambda_0}\,dw_A\wedge d\bar{w}_A &= -2\Im\left(\sum_j\oint_{A_j}dw_A\oint_{B_j}d\bar{w}_A\right)\\
&=8\pi^2\ell^i\left(\Im\tau\right)_{ij}\bar{\ell}^j\\
&=\frac{1}{2}\overline{\left(K_{im}g_m\right)}\left(\Im\tau^{-1}\right)^{ij}\left(K_{jp}g_p\right)\\
&=\frac{1}{2}\overline{\left(K_{im}g_m\sigma^{ir}\right)}\overline{\left(\frac{\partial u_r}{\partial a^k}\right)}\left(\Im\tau^{-1}\right)^{k\ell}\left(\frac{\partial u_s}{\partial a^{\ell}}\right)\left(\sigma^{sj}K_{jp}g_p\right),
\label{wasqres}\end{split}\end{equation}
where we have used the relation \eqref{sigmadef}.  To establish that $\int_{\Sigma}^{\Lambda_0}\,|dw|^2$ is proportional to \eqref{SWpot}, it remains only to show that $\sigma^{sj}K_{jp}g_p$ is equivalent to $\partial_{u_s}W_{\text{eff}}(u_q)$.  To proceed, we relate the $B$-periods $K_{im}$ to residues involving $d\omega_i$ by using the identity
\begin{equation}\begin{split}0&= \int_{\Sigma}\,d\omega_i\wedge d\Omega_m\\
&=\sum_j\left[\oint_{A_j}d\omega_i\,\oint_{B_j}d\Omega_m-\oint_{A_j}d\Omega_m\,\oint_{B_j}d\omega_i\right]+\sum_{p=\infty_+,\infty_-}\oint_{p}\left(\Omega_m\,d\omega_i\right)\\
&= K_{im}+\sum_{p=\infty_+,\infty_-}\oint_{p}\left(\Omega_m\,d\omega_i\right).\end{split}\end{equation}
The residues appearing in this expression can be simplified even further using the asymptotic behavior of the $d\Omega_m$ \eqref{Omegaasymp}{\footnote{Note that contributions from subleading behavior of $\Omega_m$ clearly vanish because $d\omega_i$ also vanishes at $\infty$.}}
\begin{equation}\oint_{\infty_{\pm}}\left(\Omega_m\,d\omega_i\right)=\oint_{\infty}\,v^m\,d\omega_i.\end{equation}
This means that
\begin{equation}K_{im}=-2\oint_{\infty}v^md\omega_i\end{equation}
and hence
\begin{equation}
\begin{split}-\frac{1}{2}\sigma^{sj}K_{jm}g_m&=\oint_{\infty}g_mv^m\sigma^{sj}d\omega_j\\
&=\oint_{\infty}g_mv^m d\eta^s\\
&=\frac{\partial}{\partial u_s}\oint_{\infty}g_mv^{m}s\,dv\\
&=-\frac{\partial}{\partial u_s}\oint_{\infty}\frac{g_mv^{m+1}}{m+1}\,ds\\
&=-\frac{\partial}{\partial u_s}\left\langle \frac{g_mv^{m+1}}{m+1}\right\rangle\\
&=-\frac{\partial}{\partial u_s}W_{\text{eff}}(u_q).
\end{split}
\end{equation}
From this we see that, up to an overall constant,
$\int^{\Lambda_0}_{\Sigma}\,|dw|^2$ is precisely the scalar potential
\eqref{SWpot} that controls the vacuum structure in the Seiberg-Witten
regime.  Namely, M-theory correctly captures physics of perturbed
Seiberg-Witten theory, because for each (off-shell) configuration in
gauge theory there is an M5 curve of the form \eqref{dw},
\eqref{dwH_dwA} which has precisely the same energy.
In principle, we can be more specific about our
particular cutoff scheme as in section \ref{subsubsec:regularization} by
determining the moduli-dependence of $\Lambda_0$ required to ensure that
$\int_{\Sigma}^{\Lambda_0}\,dw\wedge d\bar w$ is a constant.  Because
the potential \eqref{SWpot} exhibits no explicit dependence on the
cutoff scale $\Lambda_0$, though, we are precisely in the situation of
\cite{deBoer:1998by} and this is completely unnecessary.

\subsection{Implications for Exact M5 Configurations}\label{subsec:implications}

While we have seen that the potential \eqref{SWpot} arises naturally
from the M-theory framework, let us digress for a moment to discuss the
implications of this result.  Equations of motion that follow from the
expanded action \eqref{M5actSWexp} combined with the constraints
\eqref{SWbdryconds}, \eqref{SWwper} and boundary conditions \eqref{SWwvbdry} imply not
only that the moduli sit at critical points of the Seiberg-Witten
potential \eqref{SWpot} but also lead to a specific M5 embedding with
holomorphic $s,v$ and harmonic $w$.  As such, any exact solution to the
full M5 equations of motion which also satisfies \eqref{SWbdryconds}, \eqref{SWwper}, and
\eqref{SWwvbdry} must reduce to an embedding of precisely this type when
$w$ is scaled to be sufficiently small.  Among other things, this
provides another way of seeing that the M-theory realization correctly
captures supersymmetric vacua of the Yang-Mills theory.

The fact that the off-shell potential of M-theory agrees with that of
gauge theory means that even nonsupersymmetric vacua must be captured by
M-theory, not just the supersymmetric ones.  However, there is some
tension between this claim and the known facts about M5 curves.  
In particular, as first pointed out in \cite{Bena:2006rg}, holomorphic
boundary conditions of the sort \eqref{SWwvbdry} generically preclude
the existence of any exact solutions other than the supersymmetric
ones{\footnote{We will see this later in the context of specific
examples when discussing exact M5 embeddings.}}.  In that case, what are
we to make of the approximate nonsupersymmetric embeddings that follow
from \eqref{M5actSWexp}?

A crucial point is that in order to make sense of \eqref{M5actSWexp} and extract from it the Seiberg-Witten potential \eqref{SWpot}, we had to regulate it by introducing a cutoff scale $\Lambda_0$.  As such, the regulated action does not capture the behavior of the full M5 in any sense; rather, it describes local fluctuations of the M5 within a finite volume region that does not extend all the way to $\infty$.  
Approximate solutions obtained from the regulated action can thus only hope to be reliably obtained from exact ones for which we impose the condition \eqref{SWwvbdry} on some surface along $v\sim\Lambda_0$ in the interior.  Extending out toward $\infty$, the exact embeddings may exhibit starkly different behavior, even becoming wildly nonholomorphic at $\infty$.

This should not be completely unexpected because sorts of stringy
realizations that we are studying can only admit Yang-Mills descriptions at
sufficiently low energies.  Taking the NS5/D4 configurations in their
entirety corresponds to providing a specific UV completion to this
description which differs quite significantly from that of an
asymptotically free gauge theory.  To make any connection with gauge
theory, then, we must impose a cutoff scale and specify the bare
coupling constants at that scale.  In studying the regulated action
\eqref{M5actSWexp} and imposing \eqref{SWwvbdry} on the cutoff surface,
we are doing precisely that.  When we do this, however, all of the usual
caveats of effective field theory apply.  In particular, as emphasized
in \cite{ourpaper}, this means that the stringy description only
captures aspects of the Yang-Mills physics that do not explicitly depend
on the cutoff scale, $\Lambda_0$.  Note that, while the $\Lambda_0$-independence of the regulated M5 action suggests that it may accurately capture the decay of a nonsupersymmetric configuration into a supersymmetric one, one of course has the usual runaway beyond the cutoff scale associated to the difference in boundary conditions at $\infty$.

\section{Metastable Vacua in Perturbed Seiberg-Witten Theories}\label{sec:OOPP}

In the previous sections we showed how M5-branes capture the physics of
${\cal N}=2$ supersymmetric gauge theories perturbed by a small
superpotential. Theories of this type have been studied recently
\cite{Ooguri:2007iu, Pastras:2007qr} and it was found that they allow
supersymmetry breaking metastable vacua (OOPP vacua).
Because we have already demonstrated that the full gauge theory
potential in the Seiberg-Witten regime can be reproduced from off-shell M5-brane
configurations when $dw$ is sufficiently small, those OOPP vacua are guaranteed to
correspond to certain nonholomorphic M5-brane configurations of the form
\eqref{SWcurvess} and \eqref{dw} which approximately solve the equations of motion.
After quickly reviewing the results of \cite{Ooguri:2007iu,
Pastras:2007qr}, in the next section we will proceed to find such
explicit M5-brane curves corresponding to the OOPP metastable vacua,
concentrating for simplicity on the case of $SU(2)$ ${\cal N}=2$ gauge
theory.

\subsection{Review of OOPP}

If we perturb an ${\cal N}=2$ gauge theory by a small superpotential,
then to lowest order in the perturbation, the resulting scalar potential
is exactly computable. If $u^i$ are coordinates on the Coulomb branch
with K\"ahler metric $g_{i\overline{\jmath}}$, then the addition of the
superpotential $W(u^i)$ generates a scalar potential equal to:
\begin{equation}
  V(u^i) = g^{i \overline{\jmath}} \partial_i W \overline{\partial_{\overline{\jmath}}W},
\label{scalarpotentiala}
\end{equation}
which can also be written as \eqref{SWpot} in special coordinates.

It was already suggested in \cite{Intriligator:2006dd} that there might
be appropriate choices of the superpotential for which the scalar
potential \eqref{scalarpotentiala} has non-supersymmetric local
minima. In the same paper the simplest case of ${\cal N}=2$ $SU(2)$
perturbed by a quadratic superpotential $W = g \tr(\Phi^2)=g u$ for
the adjoint scalar field was considered. It was shown that in this case
the perturbation does not generate any local nonsupersymmetric
minima. The only critical points of the resulting potential are the two
supersymmetric vacua and a saddle point at $u=0$.

It was then discovered by \cite{Ooguri:2007iu, Pastras:2007qr} that
superpotentials of higher order can indeed generate metastable points.
More specifically \cite{Ooguri:2007iu} showed that, for a generic choice
of a point on the Coulomb branch of any ${\cal N}=2$ theory, it is
possible to find a superpotential perturbation which generates a
metastable vacuum at that point. As we explain below, this possibility
is based on the fact that the sectional curvature of the K\"ahler metric
on the Coulomb branch of any ${\cal N}=2$ supersymmetric gauge theory is
positive semi-definite. This follows directly from the fact that the
K\"ahler metric in (rigid) special coordinates is the imaginary part of
a holomorphic function:
\begin{equation}
  g_{i\overline{\jmath}} = \Im {\partial^2 {\cal F}(a) \over \partial a^i
\partial a^j},
\end{equation}
where ${\cal F}(a)$ is the holomorphic prepotential in terms of special 
coordinates $a^i$ on the Coulomb branch.

Following \cite{Ooguri:2007iu}, let us quickly review how one can find
the appropriate superpotential perturbation. Consider a point $p$ on the
Coulomb branch ${\cal M}$ of an ${\cal N}=2$ gauge theory, at which we
want to generate the metastable vacuum. Let $u^i$ be complex coordinates
on the moduli space near the point $p$.  We introduce K\"ahler normal
coordinates \footnote{The familiar Riemann normal coordinates on a
general K\"ahler manifold are not holomorphic. For this reason, it is
more useful to introduce the holomorphic K\"ahler normal coordinates,
which are naturally adapted to the complex structure of the manifold
\cite{AlvarezGaume:1981hn, Higashijima:2000wz}.}  ${z^i}$ around $p$,
defined by the expansion:
\begin{equation}
  z^i = u'^i + {1\over 2}\tilde{\Gamma}^{i}_{jk} u'^j u'^k +{1\over 6} 
\tilde{g}^{i\overline{m}}\partial_l(\tilde{g}_{n\overline{m}}\tilde{\Gamma}^n_{jk})u'^ju'^ku'^l,
\end{equation}
where $u'=u-u(p)$ and $\,\tilde{\,}\,$ means evaluation at $p$.

In these coordinates the metric takes the form:
\begin{equation}
  g_{i\overline{\jmath}}(z,\overline{z}) = \tilde{g}_{i\overline{\jmath}} + \tilde{R}_{i\overline{\jmath}
k\overline{l}} z^k \overline{z}^{\overline{l}} + {\cal O}(z^3).
\label{KNCexp}
\end{equation}
We choose the superpotential:
\begin{equation}
  W = k_i z^i\label{KNCsup}
\end{equation}
and find that the scalar potential around $p$ has the expansion:
\begin{equation}
  V = \tilde{g}^{i\overline{\jmath}} \partial_i W 
\overline{\partial_{\overline{\jmath}}W}
= \tilde{g}^{i \overline{\jmath}} k_i \overline{k}_{\overline{\jmath}} + k_i \overline{k}_
{\overline{\jmath}} \tilde{R}^{i\overline{\jmath}}{}_{k\overline{l}}z^k \overline{z}^
{\overline{l}} +{\cal O}(z^3).
\end{equation}
As shown in \cite{Ooguri:2007iu}, the quadratic term is positive
definite at generic points on the moduli space, therefore the vacuum at
$z^i=0$ is naturally metastable.  Because local stability depends only
on the leading terms in the expansion \eqref{KNCexp}, however, the
superpotential \eqref{KNCsup} is typically truncated to a polynomial of
finite degree.  This truncation is important to ensure that the
superpotential is well-defined on the moduli space because the K\"ahler
normal coordinates are in fact linear combinations of electric and
magnetic FI parameters \cite{MOOP}, which suffer from
monodromies{\footnote{In fact, the truncation is also important for
supersymmetry-breaking because the theory with full superpotential
\eqref{KNCsup} can realize a non-manifest ${\cal{N}}=1$ supersymmetry
which is then preserved at the $z^i=0$ vacuum
\cite{Antoniadis:1995vb,MOOP}.}}

\subsection{The Case of ${\cal N} = 2$ $SU(2)$ Gauge Theory}

In the case of $SU(2)$ the moduli space ${\cal M}$ can be parametrized by
the gauge invariant quantity:
\begin{equation}
u_2=\frac{1}{2}\left\langle\tr\Phi^2\right\rangle,
\end{equation}
which is a good global complex coordinate on ${\cal M}$. According to our previous 
discussion, the general form of the required superpotential to generate
a metastable vacuum at any point $p\in{\cal M}$ is:
\begin{equation}
  W(u) = g'(\beta' u_2^3 + \gamma'u_2^2 + u_2),
\label{multitr}
\end{equation}
where the coefficients $g',\beta',\gamma'$ depend on the choice of $p$ and 
can be easily computed from 
the K\"ahler normal coordinate expansion
around $p$. As written,  this is a multi-trace superpotential. This
is not very convenient, because in the M-theory constructions the information
of the superpotential is introduced by starting with the tree level
superpotential in a  single trace representation:
\begin{equation}
  W(\Phi) = g \left({\beta \over 6} \tr\Phi^6+ {\gamma\over 4}\tr\Phi^4
+{1\over 2} \tr\Phi^2 \right)
\label{singltr}
\end{equation}
and imposing the boundary conditions for the M5 brane on the $w$-$v$ plane:
\begin{equation}
  w(v) \sim \pm W'(v)\sim \pm g(\beta v^5 + \gamma v^3 + v).
\label{holobdcd}
\end{equation}
So we need to rewrite \eqref{multitr} in the form \eqref{singltr}.
Of course for general gauge group a multi-trace superpotential cannot always
be written, using trace identities, as a sum of single traces. However this
is always possible in the case of $SU(2)$.
To find the precise relationship
between the coefficients $g',\beta',\gamma'$ and $g,\beta,\gamma$ we need
to use the relations for the chiral ring of ${\cal N}=2$ $SU(2)$  gauge
theory. Following the analysis of \cite{Cachazo:2002ry} we have:
\begin{equation}\begin{split}
   & u_4=\frac{1}{4}\left\langle\tr\Phi^4\right\rangle = {u_2^2\over 2}
 +  {\Lambda^4\over 4},\\
  & u_6 = \frac{1}{6}\left\langle\tr\Phi^6\right\rangle = {u_2^3\over 3}
  + { u_2 \Lambda^4\over 2},
\label{ringrel}
\end{split}\end{equation}
where the terms proportional to $\Lambda^4$ are the quantum 
corrections of the chiral ring due to instantons. Using \eqref{ringrel}
the multitrace superpotential \eqref{multitr} can be written as
\eqref{singltr},

\section{An Exact M5 Curve for SU(2)} \label{sec:curves}

In this section, we explicitly construct the M5 curve corresponding to
the OOPP vacua \cite{Ooguri:2007iu, Pastras:2007qr} in gauge theory, for
the special case of $SU(2)$ gauge group.  In section \ref{sec:SW}, we
worked in a regime where the expansion \eqref{M5actSWexp} in small $dw$ was valid and used this to obtain a correspondence between M5-brane curves and gauge theory states.
We could restrict ourselves to the same regime, where the problem of
finding a minimal-area M5 curve is tantamount to minimizing the scalar
potential \eqref{SWpot}, and obtain the M5 curve of the form \eqref{dw}
and \eqref{dwH_dwA} which correspond to the OOPP vacua.
However, in this section, we will endeavor to find the \emph{exact} M5
curve without any approximation by attacking the honest
minimal-area problem.

Finding minimal-area M5 curves in such general cases is complicated but,
at the same time, illuminates the point raised in section
\ref{subsec:implications}.
As pointed out in \cite{Bena:2006rg}, if we look for M5 curves with
holomorphic boundary conditions \eqref{holobdcd} at infinity, all we can
have are holomorphic curves, corresponding to supersymmetric vacua.
Therefore, if we want nonholomorphic curves, corresponding to
nonsupersymmetric vacua, we are forced to consider \emph{nonholomorphic}
boundary conditions.
On the other hand, however, in section \ref{sec:SW} we saw that we can
realize both supersymmetric and nonsupersymmetric vacua using M5 curves
in the Seiberg-Witten regime with the \emph{holomorphic} boundary
condition such as \eqref{holobdcd}.
For things to be consistent, it must be that, when the ``bending'' along
$w$ becomes small, the nonholomorphicity of the M5 curve at infinity
becomes very small and consequently one can make the holomorphic
condition \eqref{holobdcd} hold to arbitrary precision on a boundary
surface at $\Lambda_0$ in the interior {\footnote{That one cannot simply
take the boundary surface to $\infty$ in these solutions reflects an
order of limits issue in this problem.  In particular, we will see in
this example that the parameter regime for which the expansion
\eqref{M5actSWexp} remains valid, allowing us to reproduce the
Seiberg-Witten potential \eqref{SWpot}, is one which requires $g$ in
\eqref{holobdcd} to approach zero as $\Lambda_0$ is taken to
$\infty$.}}.

In this section, we will present exact solutions to the minimal area
equations with precisely this feature.  This will allow us to accurately
define a limit in which \eqref{M5actSWexp} can be trusted and to
demonstrate that, in this limit, the exact solutions reduce to
approximate ones which solve the corresponding equations of motion.  In
particular, their moduli sit precisely at critical points of the
Seiberg-Witten potential \eqref{SWpot}.

As in \cite{ourpaper}, nonholomorphic M5 embeddings are
most easily described using a parametric framework.  As such, we shall
begin by reviewing the parametric description of the M5 configuration
realizing pure $SU(2)$ ${\cal{N}}=2$ gauge theory.  We will then turn on
a superpotential of the sort \eqref{singltr} by suitably ``bending'' the
M5 brane in the $w$ and $v$ directions.  Within this setup, we will look
for minimal area nonholomorphic solutions corresponding to the OOPP
vacua of the gauge theory.

\subsection{The Minimal Area Problem}

First, however, let us review a few technical results that will be
useful for our parametric representation of M5 configurations.  The
mathematical problem that we have to solve is to find a minimal-area
embedding of a Riemann surface $\Sigma$ in the space ${\mathbb
R}^5\times S^1$, which is parametrized by the complex coordinates
$w,v,s$. Because of the identification $s\sim s+ 2\pi i$, it is
convenient to introduce the coordinate $t=\Lambda^4 e^{-s}$ as in
\eqref{tdef}.  In the case of supersymmetric $M5$ configurations it is
easy to describe the embedding by polynomial equations such as
\eqref{SWcurves} and \eqref{wvcurves}, which more generally take the
form:
\begin{equation}\begin{split}
  &F_1(v,t) = 0,\\
  &F_2 (v,w) = 0.
\end{split}\end{equation}
Such a representation of the surface is not very convenient when we want
to consider non-holomorphic embeddings, as we have found \cite{ourpaper}
that the parametric description is often more suitable for this
purpose. For this, we consider a Riemann surface $\Sigma$ with
holomorphic coordinate $z$ on its ``worldsheet'', and describe the
embedding of the M5 brane by the functions:
\begin{equation}
v(z,\overline{z}) , \qquad  w(z,\overline{z}),\qquad  s(z,\overline{z}).
\label{embf}
\end{equation}
Since the $M5$ brane is non-compact, the embedding functions will have
poles on certain points of the Riemann surface corresponding, in the
type IIA picture, to the infinities of the NS5-branes. This means that,
more precisely, $\Sigma$ is a punctured Riemann surface.

We now turn to the equations of motion for the embedding functions \eqref{embf}.
In general if we have
a two dimensional surface $X^\mu(\sigma^a)$ embedded in an ambient space
of metric $G_{\mu\nu}$, then the induced metric on the surface is:
\begin{equation}
  g_{ab} = G_{\mu\nu}\partial_a X^\mu \partial_b X^\nu
\end{equation}
and the area is given by the expression:
\begin{equation}
  A = \int_\Sigma \sqrt{\det(g)}
\label{NGACT}
\end{equation}
As we know very well from the case of the bosonic string, it is very
useful to pick the coordinates $\sigma^i$ in such a way that the induced
metric is in conformal gauge. We use the complex variable $z$ for this
class of coordinates.  In this gauge the equations of motion from
varying the ``Nambu-Goto'' action \eqref{NGACT} become equivalent to two
conditions: the first is that the embedding functions
$X^\mu(z,\overline{z})$ must be harmonic and the second that the
Virasoro constraint must be satisfied:
\begin{equation}
  G_{\mu\nu} {\partial X^\mu \over \partial z} {\partial X^\nu \over \partial z
} =0.\label{Vconst}
\end{equation}
In our case, the situation is particularly simple because the background
metric $G_{\mu\nu}$ is flat and harmonic functions in two-dimensions are
simply sums of holomorphic and antiholomorphic ones.  As such, our
embedding functions take the form
\begin{equation}\begin{split}
   v(z,\overline{z}) &= v_H(z) + \overline{v_A(z)},\\
   w(z,\overline{z}) &= w_H(z) + \overline{w_A(z)},\\  
   s(z,\overline{z}) &= s_H(z) + \overline{s_A(z)}
\label{decomp}
\end{split}\end{equation}
for holomorphic $v_{H/A}^{}$, $w_{H/A}^{}$, $s_{H/A}^{}${\footnote{More precisely 
meromorphic since they can have poles at the punctures of the Riemann 
surface.} and the constraint \eqref{Vconst} becomes
\begin{equation}
  \partial v_H \partial v_A +  \partial w_H \partial w_A +
   R_{10}^2\partial s_H \partial s_A =0.
\label{vircon}
\end{equation}
Note that although this is a nonlinear condition, it is nevertheless a
\emph{holomorphic} one, a property which makes it considerably easier to
solve.  For any set of holomorphic functions $v_{H/A}^{}$, $w_{H/A}^{}$,
$s_{H/A}^{}$ satisfying \eqref{vircon}, the embedding given by
\eqref{decomp} extremizes\footnote{Checking
whether this extremum is truly a minimum (locally stable)
is extremely difficult. To answer this question one has to consider the
second variation of the area functional \eqref{NGACT} around the local
extremum, and check that there are no deformations that locally decrease
the area. The authors are not aware of any general method of checking
local stability for an arbitrary embedding.  In the simplifying limits
of sections \ref{sec:DV} and \ref{sec:SW}, though, this question is
easier to address and translates into the matter of local stability in
the corresponding field theory potentials.} the area
\eqref{NGACT}.

Finally, to specify the lift of a given NS5/D4 configuration, we must
impose suitable conditions on our embedding functions as discussed in
section \ref{sec:IIASW}.  In particular, for $s$ we must fix the period
integrals \eqref{SWbdryconds} about the $\hat{A}$- and $\hat{B}$-cycles
of figure \ref{vplane}
\begin{equation}\frac{1}{2\pi i}\oint_{\hat{A}_j}\,ds=N_j\qquad\text{and}\qquad \frac{1}{2\pi i}\oint_{\hat{B}_j}\,ds=-\alpha_j.\label{SWbdrycondsb}\end{equation}
As for $w$ and $v$, they take values in $\mathbb{R}^2$ and hence $dw$
and $dv$ must have vanishing periods around all compact cycles in the
geometry.  Phrased in terms of the $\hat{A}$- and $\hat{B}$-cycles of
figure \ref{vplane}, this condition amounts to
\begin{equation}\oint_{\hat{A}_j}dw=0\text{ for all }j,\qquad \oint_{\hat{B}_i-\hat{B}_j}dw=0\text{ for all }i,j\label{SWwperb}\end{equation}
and similar equations for $dv$.  Their asymptotic behavior at $\infty$
is then fixed by the superpotential{\footnote{Of course, as discussed
earlier in this section we will have to suitably relax this constraint
later when looking for nonsupersymmetric solutions.}}
\begin{equation}w\sim \pm W'(v).\label{SWwv}\end{equation}
In terms of the embedding functions $v(z,\bar{z})$ and $w(z,\bar{z})$,
this condition imposes nontrivial relations between their pole
structures at the punctures.

To summarize, the problem we have to solve is to find meromorphic
functions $v_{H/A}^{}$, $w_{H/A}^{}$, $s_{H/A}^{}$ on a (punctured) Riemann
surface $\Sigma$, satisfying the Virasoro constraint \eqref{vircon} and
such that the embedding functions \eqref{decomp} have the monodromies
\eqref{SWbdrycondsb}, \eqref{SWwperb}, and correct boundary conditions
\eqref{SWwv}.

\subsection{Harmonic Functions on the Torus}
\label{subsec:harmfuncsT2}

We now specialize to the case of $SU(2)$, where our setup is
particularly simple.  The M5 lift, $\Sigma$, is a genus one curve with
two punctures corresponding to the points at $\infty$ on the two
NS5-branes.  Consequently, the embedding is described by meromorphic
functions on a torus.  In this section, we will introduce a convenient
collection of such meromorphic functions and review their basic
properties.

To parametrize the M5 curve, we use the complex $z$-plane subject to the identifications
\begin{equation}
z \sim z+1 \sim z+\tau  .
\end{equation}
We also have to specify the two marked points $a_1,a_2$ which correspond
to the points at $\infty$ of the NS5-branes.  Notice that we can always
use the conformal killing vectors of the torus to set one of the
punctures to any desired point on the torus so that only the difference
$a=a_2-a_1$ has an invariant meaning.  In figure \ref{M5params}, we
depict both the $v$ and $z$ parametrizations of our curve along with the
corresponding realizations of the $\hat{A}$- and $\hat{B}$-cycles.

\begin{figure}
\begin{center}
\subfigure[Parametrization of the genus 1 M5 embedding as a double cover of the $v$-plane with $\hat{A}$ and $\hat{B}$ cycles indicated]
{\epsfig{file=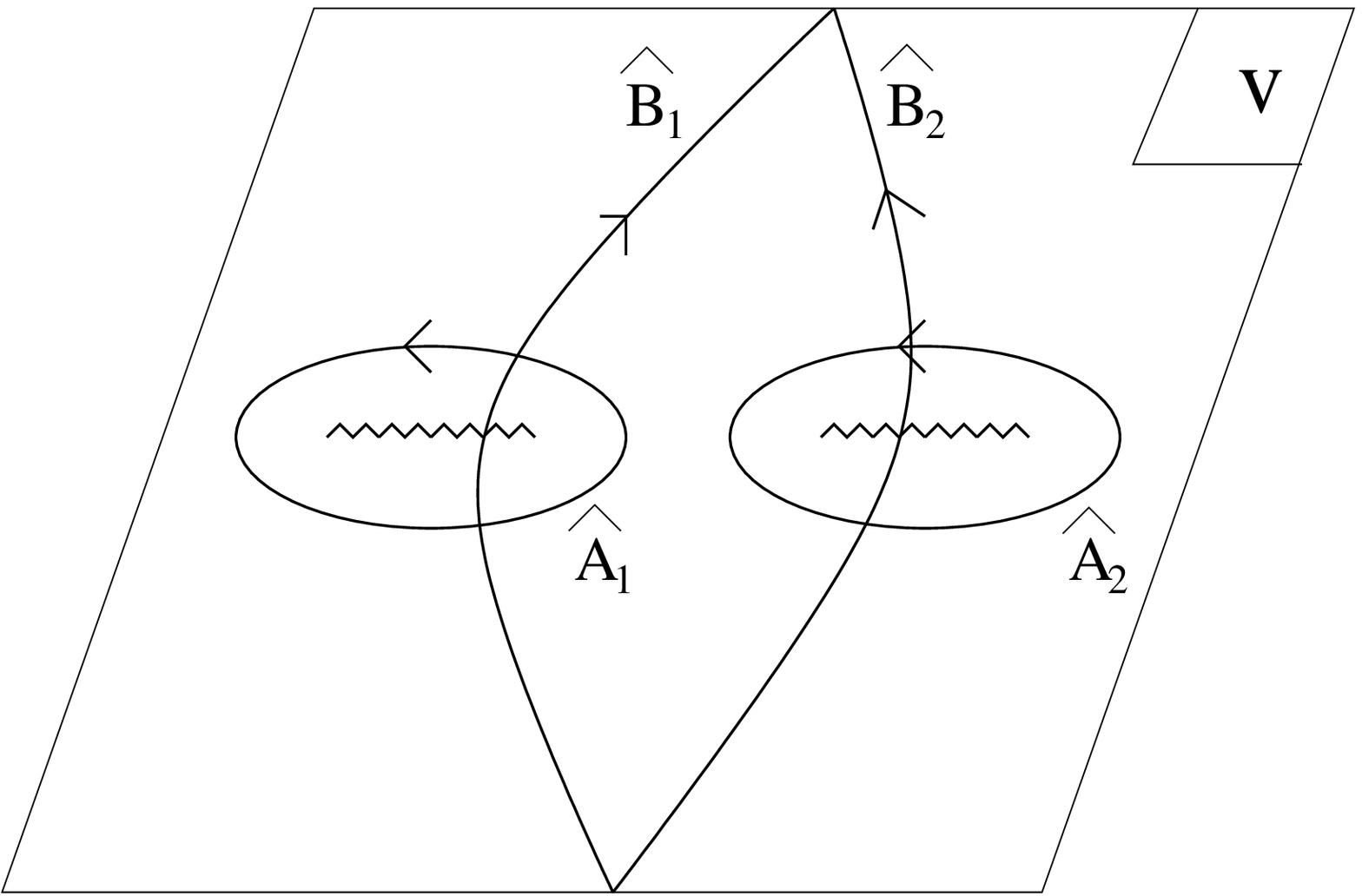,width=0.4\textwidth}\label{cubsupex}}
\subfigure[Parametrization of the genus 1 M5 embedding via the fundamental parallelogram on the $z$ plane with $\hat{A}$ and $\hat{B}$ cycles indicated]
{\epsfig{file=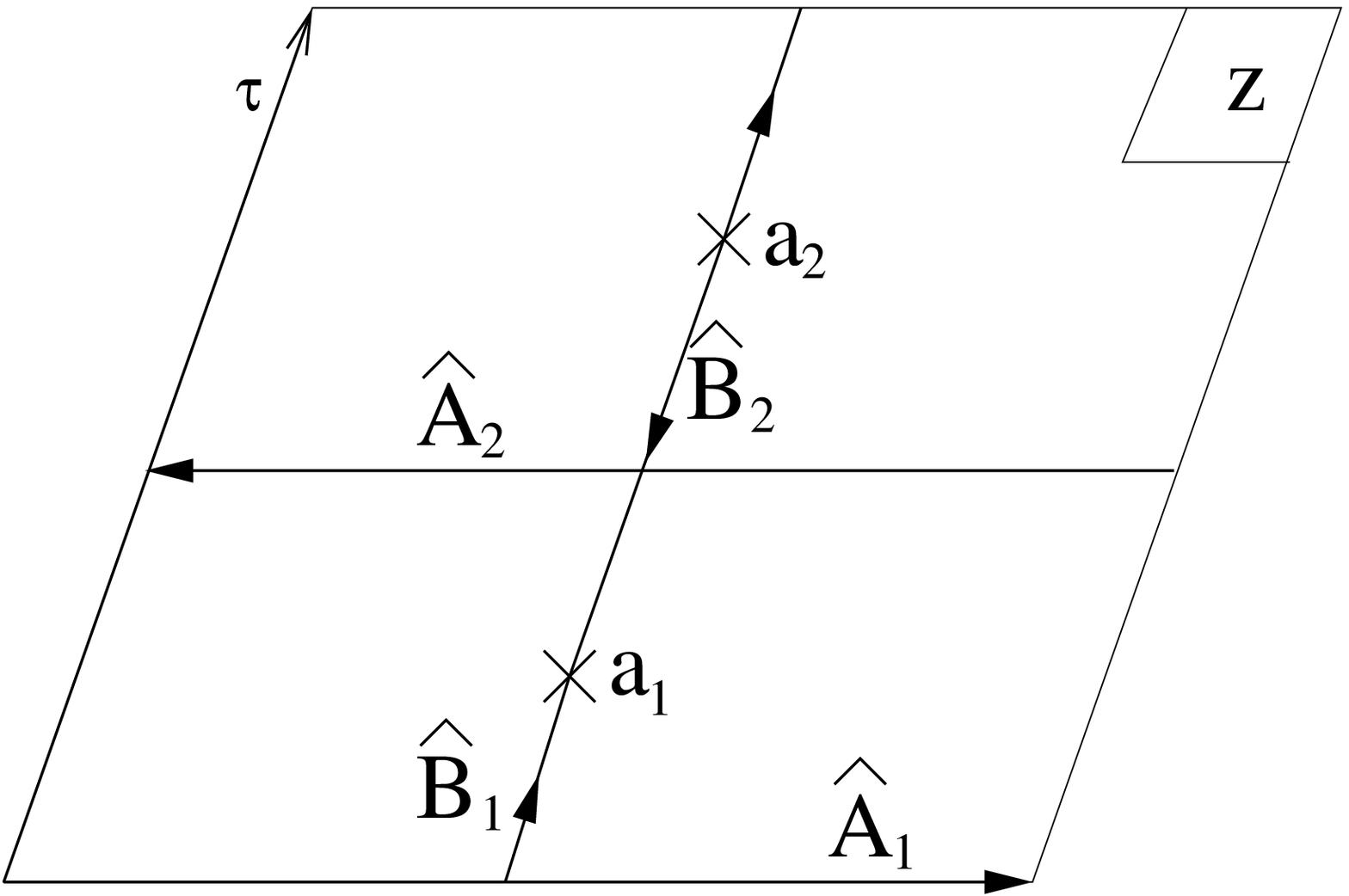,width=0.4\textwidth}\label{zplane}}
\caption{Parametrizations of the genus 1 M5 embedding}
\label{M5params}
\end{center}
\end{figure}

As we saw in the previous section, the minimal area equations imply that
the embedding coordinates of the $M5$ brane must be harmonic functions
of the coordinate $z$, possibly with singularities at the punctures,
with fixed monodromies around the compact cycles of the surface.  As we
explain in Appendix \ref{app:RiemannS}, it is possible to write the most
general harmonic 1-form $f$ on a punctured Riemann surface in terms of
the standard differentials, namely, holomorphic differentials (a.k.a.\
meromorphic differentials of the first kind) $\omega_i$, meromorphic
differentials of the second kind $d\Omega_n^P$, and meromorphic
differentials of the third kind $d\Omega_0^{P,P'}$.

In the
case of our genus one surface, there is only one holomorphic
differential, namely
\begin{equation}\omega = 2\pi i\,dz,\label{holodiff}\end{equation}
whose periods on the standard compact $A$ and $B$ cycles are:
\begin{equation}
\frac{1}{2\pi i}\int_{\hat{A}_1} \omega = 1,\qquad \frac{1}{2\pi i}\int_{\hat{B}_1-\hat{B}_2} \omega = \tau.
\end{equation}
To explicitly construct meromorphic differentials with desired poles at
the punctures, we start with the basic theta function:
\begin{equation}
\theta(z)=\sum_{n=-\infty}^{\infty}e^{i\pi n^2\tau+2\pi inz},
\label{thetadef}
\end{equation}
and use it to define the (quasi-)elliptic functions:
\begin{equation}
F(z)=\ln\theta(z-\tilde{\tau}),\qquad\qquad
\tilde{\tau}\equiv \frac{1}{2}(\tau+1),
\label{Fdef}
\end{equation}
\begin{equation}
F_i^{(n)}=\left(\frac{\partial}{\partial z}\right)^{n}F(z-a_i).
\label{Fderivs}
\end{equation}
Because $F(z)\sim\ln z$ near $z=0$, the function $F_i^{(n)}$ introduces
an $n$th order pole at the point $a_i$. For $n\ge 2$ these functions are
elliptic, while for $n=0,1$ they have the following monodromies:
\begin{equation}\begin{split}
F_i(z+1)&=F_i(z),\\
F_i(z+\tau)&=F_i(z)+i\pi - 2\pi i(z-a_i),\\
F_i^{(1)}(z+1)&=F_i^{(1)}(z),\\
 F_i^{(1)}(z+\tau)&=F_i^{(1)}(z)-2\pi i.
\label{Fmonods}
\end{split}\end{equation}
Useful information about these elliptic functions, including their
relation to Weierstrass functions, can be found in the appendices of
\cite{ourpaper}.

Before we close this section, let us summarize the relation between
these elliptic functions and the standard meromorphic differentials.  On
a genus one surface the only holomorphic differential is $\omega$
\eqref{holodiff}.  One basis for meromorphic differentials of the second
kind having a $(n+1)$th order pole at the point $a_i$ are
\begin{equation}
  d\Omega^{a_i}_n =  {(-1)^n\over (n-1)!}F_i^{(n+1)} dz,\qquad n\geq 1.
\end{equation}
Finally the meromorphic differential of the third kind with simple poles
at two points $a_1,a_2$ and opposite residues is:
\begin{equation}
  d\Omega_0^{a_1,a_2} = (F_1^{(1)} - F_2^{(1)} + i \pi) dz.
\end{equation}

\subsection{The ${\cal N}=2$ Curve in Parametric Representation}\label{subsec:SWparam}

To see how we can use the elliptic functions to set up the parametric
representation of the surface, let us try to write the ${\cal N}=2$
curve which describes the Coulomb branch of the gauge theory without
superpotential in this formalism{\footnote{Such a parametric representation is also described in Appendix B of \cite{Mazzucato:2007ah}.}}.  As usual, the IIA configuration
consists of parallel NS5's with two D4's suspended in between, with its
M-theory lift given by a single M5 extended on the genus one holomorphic
Riemann surface corresponding to the $SU(2)$ Seiberg-Witten curve
\begin{equation}t^2-2P_2(v)t+\Lambda^4=0,\label{embedn2}\end{equation}
where
\begin{equation}P_2(v)=v^2-u_2.\end{equation}

Now we want to describe a parametric representation of this curve using
the elliptic functions introduced in the previous section. We start with
the embedding $v(z)$.  This function must have first order poles at the
two punctures of the torus, which correspond to the infinite regions of
the NS5-branes.  We are thus led to the expression:
\begin{equation}
  v(z) = b_2 \left( F_1^{(1)} - F_2^{(1)} + i\pi\label{vSW}\right),
\end{equation}
where the $i \pi$ shift has been added for convenience{\footnote{Without this shift, we would obtain a generic $U(2)$ curve with nonzero $u_1$.  Including this shift will set $u_1=0$, leaving us with the $SU(2)$ curve.}} and $b_2$ is a constant whose value will be fixed later.

The function $v(z)$ is holomorphic and single-valued as it should be.
On the other hand, we know from \eqref{SWbdryconds} that $s(z)$ is
multivalued, having winding number 1 around each of the $\hat{A}$-cycles
of Figure \ref{M5params},
\begin{equation}\frac{1}{2\pi i}\oint_{\hat{A}_j}\,ds=1,\qquad j=1,2.\label{SWdsconst}\end{equation}
Up to a constant shift, there is a unique holomorphic $s(z)$ which satisfies these conditions
\begin{equation}s(z)=2\left(F_1-F_2+i\pi z\right).\label{sSW}\end{equation} 
We now turn to the $\hat{B}$ period-constraints \eqref{SWbdryconds}:
\begin{equation}\oint_{\hat{B}_j}\,ds=-\alpha,\qquad j=1,2.\label{Bhatconsts}\end{equation}
The condition
\begin{equation}\left(\oint_{\hat{B}_2}-\oint_{\hat{B}_1}\right)\,ds=0\quad\implies\quad s(z+\tau)=s(z)\end{equation}
is particularly simple and fixes the distance $a$ between the punctures to be
\begin{equation}a\equiv a_2-a_1=\frac{\tau}{2}.\end{equation}
The final constraint fixes $b_2$ as a function of $\tau$ and the
dynamical scale, $\Lambda$.  The simplest way to derive the correct
relation is to demonstrate, as we do in Appendix \ref{app:SWparam}, that
$v(z)$ \eqref{vSW} and $s(z)$ \eqref{sSW} satisfy the equation
\eqref{embedn2} under the identification
\begin{equation}b_2^4=\frac{\Lambda^4}{12\wp(\tau/2)^2-g_2},\label{b2Lambda}\end{equation}
where $\wp(z)$ is the Weierstrass $\wp$-function with the half-periods
$(\omega_1,\omega_2)=(1/2,\tau/2)$ and $g_2$ is one of the Weierstrass
elliptic invariants.

The result is a family of holomorphic embeddings parametrized by the
complex structure modulus, $\tau$.  As shown in Appendix
\ref{app:SWparam}, we can trade $\tau$ for the more conventional
modulus $u_2$ via
\begin{equation}u_2=3b_2^2\wp(\tau/2).\label{u2b2}\end{equation}

\subsection{Turning on a Superpotential}

Introduction of a superpotential is achieved by bending the
M5 brane in the $w$ direction through the boundary conditions
\begin{equation}
  w(v) \sim \pm W'(v)
\label{wvbcagain}
\end{equation}
as $v\rightarrow\infty$.  As we saw in section 2, it is rather simple to
find the supersymmetric configurations compatible with these boundary
conditions.  The $w$ bending \eqref{wvbcagain} can be realized by a
holomorphic embedding at discrete points on the Coulomb branch which
correspond precisely to extrema of the superpotential.  Moreover it is
possible to find these points by the factorization conditions
\eqref{facteqn},\eqref{facteqnn}, which together with \eqref{SWcurves}
elegantly give the form of the functions $v,w,s$ for the supersymmetric
embeddings.

As discussed before, when we want to consider nonholomorphic minimal
area embeddings, the situation is more complicated.  The main difficulty
is that, strictly speaking, it is not possible to find nonsupersymmetric
minimal area embeddings satisfying \eqref{wvbcagain}.  To see why, let
us return to the constraint \eqref{vircon}
\begin{equation}
\partial v_H\partial v_A+\partial w_H\partial w_A+R_{10}^2\partial s_H\partial s_A=0\label{virconn}
\end{equation}
and study the pole structure of various terms that appear.  Because the
embedding is locally one-to-one as $v\rightarrow\infty$, $v_H$ has first
order poles at the marked points.  The holomorphic function $w_H$, on
the other hand, has poles of order $n$, the degree of $W'(v)$.  Imposing
purely holomorphic boundary conditions $w(v)$ near $v\rightarrow\infty$
further implies that $v_A$ and $w_A$ have no poles at all{\footnote{Even
though this implies that any nontrivial contribution to $v_A$ or $w_A$
undergoes monodromies around the nontrivial cycles of the torus, this is
not a problem because it is only the combinations $v_H+\bar{v}_A$ and
$w_H+\bar{w}_A$ that must be well-defined.}}.  This means that the $v$
contribution to \eqref{virconn} contains poles of order 2 and lower
while the $w$ contribution contains poles of order $n+1$ and lower.
Finally, because $ds$ has at most first order poles, the $s$
contribution to \eqref{virconn} also contributes poles of order 2 and
lower.  As a result, the contribution from $w$ cannot be cancelled by
either of the others and an exact solution is not
possible{\footnote{Note that we can in principle have solutions where
$w$ is holomorphic and $s,v$ are not.  From our analysis of section
\ref{sec:SW}, though, it is clear that the solutions we are looking for
are not of this type.}}.
 
To find solutions, then, it is clear that we will have to introduce
antiholomorphic contributions to $v$ with higher order poles and even
possibly new holomorphic contributions as well.  In what sense can we
make a connection with the analysis of section \ref{sec:SW}, then?  As
discussed in section \ref{subsec:implications}, there is no reason for
us to expect that exact nonholomorphic solutions exist which satisfy the
condition \eqref{wvbcagain} at $v=\infty$.  Rather, we can only expect
that it holds at small $w$ to arbitrary precision on a boundary surface
in the interior.

That this can happen is quite easy to see directly from our analysis so far.  If we start with an exact solution and scale $w$ to be parametrically small, \eqref{virconn} suggests that the antiholomorphic contributions to $v$ will decrease even more rapidly, provided of course that the corresponding holomorphic ones remain finite.
In that sense, we may expect to find solutions which approximately
approach embeddings of the form advertised in section
\ref{subsec:implications}, namely those for which $s$ and $v$ describe
holomorphic Seiberg-Witten geometries with a harmonic embedding along
$w$.  Because the antiholomorphic terms contain poles, however, their
contribution can never be made parametrically small at $\infty$.
Rather, we must introduce a fixed cutoff surface along $v\sim \Lambda_0$
and then scale $w$ to zero while keeping $\Lambda_0$ fixed.  For
$|v|<|\Lambda_0|$, such solutions will take the form of section
\ref{subsec:implications}.  Farther toward $\infty$, however, they will
remain wildly nonholomorphic.  As a result, if we insist on finding
minimal area solutions, we will have to relax the boundary condition
\eqref{wvbcagain} slightly.

\subsection{The Exact Solution}

In this subsection, we will motivate a suitable ansatz and present a
family of exact solutions which will be relevant for the $SU(2)$ gauge
theory.  In the following subsection, we will then make the connection
to OOPP vacua more precise by studying this family of solutions in a
limit where the expansion \eqref{M5actSWexp} is justified.

We call $z$ the holomorphic coordinate on the worldsheet of the torus.
In the supersymmetric case the function $v(z,\overline{z})$ has first
order poles around the punctures. For the reasons mentioned in the last
section, we consider a more general ansatz where the function
$v(z,\overline{z})$ has also higher order and nonholomorphic poles. We
will take the ansatz for $w(z,\overline{z})$ to be dictated by the
degree of the superpotential, and include higher order and
nonholomorphic poles in $v(z,\overline{z})$ that are minimal for
satisfying the Virasoro constraint.  For $s(z,\overline{z})$, we include
only logarithmic bending.  As we discuss in Appendix \ref{app:symm}, the
form of the ansatz is also constrained by requiring the symmetries that
supersymmetric curves possess in the $SU(2)$ case.
From all these considerations, we are led to study the following ansatz:
 \begin{equation}\begin{split}
& w = A_6 \left(F_1^{(5)} + F_2^{(5)}\right)+ A_4 
\left(F_1^{(3)} + F_2^{(3)}\right)\\
&\qquad +A_2 \left(F_1^{(1)} + F_2^{(1)}+{2\pi \over \tau_2}\left[
(z-A)-(\overline{z}-\overline{A})\right]-2\pi i\right),\\
& v = b_4 \left(F_1^{(3)}- F_2^{(3)}\right) + b_2 \left(F_1^{(1)}- F_2^{(1)}
+i \pi\right)\\
&\qquad +\overline{c}_4 \left(\overline{F}_1^{(3)}- \overline{F}_2^{(3)}\right)
+ \overline{c}_2 \left(\overline{F}_1^{(1)}- \overline{F}_2^{(1)}+ i \pi
\right),\\
& s = (2 + \nu) \left(F_1 - F_2+ i \pi z\right)+ \nu \left(
\overline{F}_1 -\overline{F}_2 - i \pi \overline{z}\right),
\label{SUtwoExacta}
\end{split}\end{equation}
where
\begin{equation}A\equiv\frac{a_1+a_2}{2}\end{equation}
and
\begin{align}
 \tau=\tau_1+i\tau_2.
\end{align}
Finally we have to fix the undetermined coefficients by demanding that
our ansatz satisfies the Virasoro constraint. This leads to the
solution{\footnote{There is also an analogous solution with $c_4=0$
which is related to this one by complex conjugation of $v$.}}
\begin{equation}\begin{split}
b_4 &=0,\\
c_4 &= -{40 \pi \overline{A}_2 A_6 \over b_2 \tau_2},\\
c_2 &= -{12 \pi \overline{A}_2(A_4 + 20 A_6 \wp(\tau/2))\over b_2 \tau_2},\\
R_{10}^2\tau_2\bar{\nu}(\nu+2)&=-2\pi\bar{A}_2\left[A_2+48A_6\left(15\wp(\tau/2)^2-g_2\right)+12A_4\wp(\tau/2)\right],\\
0&=4\tau_2\left(A_4+12A_6\wp(\tau/2)\right)\left(g_2-3\wp(\tau/2)^2\right)+A_2\left[2\pi+(\wp(\tau/2)-4\eta_1)\tau_2\right].
\label{SUtwoExactb}
\end{split}\end{equation}
This describes a family of minimal area surfaces\footnote{Actually we
only know that these are surfaces with extremal area. We have not been
able to find a practical method to determine when the area is truly
minimal in generic situations.  In the OOPP limit, this question can be
answered by studying the corresponding scalar potential, which was
discussed in section \ref{sec:OOPP}.}. 

Once we go away from the small superpotential limit or the
Seiberg-Witten regime, there is some degree of arbitrariness in turning
on higher order and nonholomorphic terms in $w,v,s$.  For example,
although in the ansatz \eqref{SUtwoExacta} we required the symmetries
that supersymmetric $SU(2)$ curves possess, we could have considered an
ansatz which does not possess this symmetry and might have ended up with
a solution which is different from and presumably much more complicated
than \eqref{SUtwoExactb}.  However, the point here is to show the existence
of an exact solution which reduces to the approximate solution derived
in the last section, thus justifying the expansion \eqref{M5actSWexp}.
We will see that the simple solution \eqref{SUtwoExacta} does have such
a property, and therefore it is a sufficient solution for our purpose.

In the following section we will study the physical meaning of the
various quantities which parametrize this family.

\subsection{The Perturbed ${\cal N}=2$ Regime}
\label{subsec:pertN=2}

In this section we would like to consider a limit of the solution
\eqref{SUtwoExacta}, \eqref{SUtwoExactb} such that, within some
appropriate boundary surface, the curve takes the approximate form of a
holomorphic Seiberg-Witten geometry $s(v)$ with a harmonic embedding
along $w$.  Within this limit, we will then attempt to impose boundary
conditions of the sort \eqref{holobdcd} on that boundary surface.  This
will be self-consistent only provided $g$ is small in a sense that we
will make more precise below.  In the end, this will allow us to explicitly realize
OOPP vacua in the manner suggested by the analysis of section
\ref{sec:SW}.

\subsubsection{The OOPP Limit}

The parametric representation \eqref{vSW} and \eqref{sSW} of the $SU(2)$
Seiberg-Witten curve gives some guidance as to what our limiting curve
should look like.  In particular, it suggests that we consider a regime
in which we can simply neglect the terms proportional to $c_2$, $c_4$,
and $\nu$ in \eqref{SUtwoExacta} so that our solution takes the
approximate form
\begin{equation}
\begin{split}
w&=A_6\left(F_1^{(5)}+F_2^{(5)}\right)+A_4\left(F_1^{(3)}+F_2^{(3)}\right)\\
&\qquad
+A_2\left(F_1^{(1)}+F_2^{(1)}+\frac{2\pi}{\tau_2}\left[(z-A)-(\bar{z}-\bar{A})\right]-2\pi i\right),\\
 v&=b_2\left(F_1^{(1)}-F_2^{(1)}+i\pi\right),\\
s&=2\left(F_1-F_2+i\pi z\right). \label{OOPPsol}
\end{split}
\end{equation}
From \eqref{SUtwoExactb}, we see that this can be accomplished by
suitably scaling $A_2$, $A_4$, and $A_6$ to zero.  A parametric
separation can be achieved which results in the approximate form
\eqref{OOPPsol} because $A_2$, $A_4$, and $A_6$ are linearly related to
one another while $c_2$, $c_4$, and $\nu$ depend on them in a quadratic
manner.  We must be a bit careful, though, because the $c_2$, $c_4$, and
$\nu$ terms of \eqref{SUtwoExacta} that we want to neglect nevertheless
become arbitrarily large near the marked points, where they exhibit
divergences of varying degree.  This means that we can never take a
limit in which our solution looks everywhere like \eqref{OOPPsol}.
Rather, the best we can hope for is that our curve approximately
resembles \eqref{OOPPsol} only after the divergences have been suitably
regulated.  For this, we introduce a cutoff by removing a circle of
radius $\epsilon$ about each marked point.  Holding $\epsilon$ fixed, we
can then take $A_2$, $A_4$, and $A_6$ sufficiently small that our
solution \eqref{SUtwoExacta} approaches \eqref{OOPPsol} to arbitrary
precision inside the regulated surface.  We shall hereafter refer to this as
the ``OOPP limit''.

Precise inequalities which yield the OOPP limit are most easily
determined by studying the conditions for which \eqref{OOPPsol} itself
becomes an approximate solution to the equations of motion.  More
specifically, we want to scale $A_2$, $A_4$, and $A_6$ in such a way
that the Virasoro constraint \eqref{vircon} holds to arbitrary precision
inside the cutoff surface.  It is easy to see that, for generic moduli
$\tau$, we simply need
\begin{equation}\frac{A_6A_2}{\tau_2}\ll \epsilon^6\qquad \frac{A_4A_2}{\tau_2}\ll \epsilon^4\label{epsconds}\end{equation}
These can in turn be translated into relations involving the physical
cutoff $\Lambda_0$ by noting that, in the OOPP limit,
\begin{equation}\epsilon = \frac{b_2}{\Lambda_0}.\end{equation}
Note that applying the conditions \eqref{epsconds} to our exact solution
\eqref{SUtwoExacta}--\eqref{SUtwoExactb} leads to suppression of the
$c_2$, $c_4$, and $\nu$ terms as expected{\footnote{The $\epsilon$
suppression might seem to be larger than necessary at first glance but
we must keep in mind that it is not sufficient for the nonholomorphic
terms in \eqref{SUtwoExacta} to simply scale like a positive power of
$\epsilon$.  Rather, we want nonholomorphic corrections to the boundary
conditions at $\Lambda_0$ to become negligible and, for this, higher
order suppression of various nonholomorphic parts of the embedding can be needed.}}.

\subsubsection{The Approximate Curve and Gauge Theory Quantities}

We now turn to a study of the approximate curve \eqref{OOPPsol} to which
our exact solution \eqref{SUtwoExacta} reduces in the OOPP limit.  
More specifically, we have a family of curves characterized by five independent parameters, $A_6$, $A_4$, $A_2$, $b_2$, and $\tau$.  On this family, we are now free to specify the boundary condition \eqref{holobdcd}  
\begin{equation}w(v)\sim \pm g\left(\beta v^5+\gamma v^3 +v\right).\label{superpotgt}\end{equation}
By studying the expansion of the elliptic functions $F_i^{(n)}$, this
fixes $A_2$, $A_4$, and $A_6$ in terms of $g$, $\beta$, and $\gamma$ as
\begin{equation}\begin{split}
A_6&=g\frac{\beta b_2^5}{24},\\ A_4&=g\frac{\gamma b_2^3+5\beta
b_2^5\wp(\tau/2)}{2},\\ A_2&=g\frac{2b_2+6\gamma b_2^3\wp(\tau/2)+\beta
b_2^5\left[30\wp(\tau/2)^2-g_2\right]}{2}.
\label{Asbdrycond}\end{split}\end{equation} Furthermore, recognizing the
$v$ and $s$ embeddings as equivalent to our parametric description of
the Seiberg-Witten geometry, \eqref{vSW} and \eqref{sSW}, we can
immediately read off the relation \eqref{b2Lambda}
\begin{equation}b_2^4=\frac{\Lambda^4}{12\wp(\tau/2)^2-g_2},\label{b2tau}\end{equation}
which follows from the boundary conditions that we impose along $s$.
Already this is enough to rephrase the OOPP limit \eqref{epsconds} in
terms of quantities which enter our boundary conditions.  For example, if we suppose
that $\tau$ is generic and both $\beta\Lambda^4$ and $\gamma\Lambda^2$
are of ${\cal{O}}(1)$, we see that the approximate solution
\eqref{OOPPsol} with boundary conditions \eqref{superpotgt} can only
arise as the OOPP limit of our exact solution provided
\begin{equation}g\ll \left(\frac{\Lambda}{\Lambda_0}\right)^3.\end{equation}
Of course, we can have more complicated inequalities for non-generic
choices of parameters.  In our previous analysis we simply said that $g$
had to be ``sufficiently small'' so it is nice to see a more precise
condition arise{\footnote{Note that the appearance of $\Lambda_0$
provides yet another demonstration of the importance of regulating the
surface before imposing \eqref{superpotgt}.  If we take $\Lambda_0$ to
be strictly infinite then our reduced curve is never an approximate
solution for any nonzero $g$ no matter how small.}}.

Let us now return to the one remaining unfixed parameter, namely the
complex structure modulus $\tau$.  From our analysis of section
\ref{sec:SW}, we expect that this should be determined by the
Seiberg-Witten potential \eqref{SWpot}.  In the case at hand, however,
we note that $\tau$ is fixed by the equations of motion of the exact
solution from which our approximate curve \eqref{OOPPsol}
descends. 
Indeed, looking at \eqref{SUtwoExactb}, the first four equations can be
thought of as fixing the parameters $b_4$, $c_4$, $c_2$, and $\nu$,
which are all negligible in the OOPP limit.  The last equation, however,
yields a nontrivial relation which must be satisfied among the four
remaining parameters $A_2$, $A_4$, $A_6$, and $\tau$ of the approximate
solution \eqref{OOPPsol}.  Plugging in the values \eqref{Asbdrycond}, this equation becomes
\begin{equation}\begin{split}0&=2b_2g\left(g_2-3\wp(\tau/2)^2\right)\left(6\beta b_2^4\wp(\tau/2)+\gamma b_2^2\right)\\
&\qquad\qquad+\frac{b_2g}{2}\left[2+\beta b_2^4\left(30\wp(\tau/2)^2-g_2\right)+6\gamma b_2^2\wp(\tau/2)\right]\left(\wp(\tau/2)-4\eta_1+\frac{2\pi}{\tau_2}\right).\label{compA2eqn}\end{split}\end{equation}
To keep notation from getting out of control, we do not plug in the result \eqref{b2tau} for $b_2$.  Nevertheless, we should remember that $b_2$ is itself a function of $\tau$.

In the end, we have found that, in the OOPP limit, our exact solution
reduces to a holomorphic Seiberg-Witten geometry with harmonic $w$
embedding and complex structure modulus $\tau$ determined by solving the
complicated equation \eqref{compA2eqn}.

\subsubsection{Explicit Connection to Seiberg-Witten Potential}

From the general discussion in section \ref{sec:SW}, we expect that the equation \eqref{compA2eqn} is nothing other than the condition for $\tau$ to be a critical point of the Seiberg-Witten potential associated to the superpotential
\begin{equation}W(\Phi)=g\left(\frac{\beta \Phi^6}{6}+\frac{\gamma \Phi^4}{4}+\frac{\Phi^2}{2}\right).\end{equation}
We will now come full circle and demonstrate this explicitly by studying the corresponding Seiberg-Witten potential \eqref{SWpot} given by
\begin{equation}V_{SW}=\frac{1}{\tau_2}\left|\partial_{u_2}W_{\text{eff}}(u_2)\right|^2\left|\frac{\partial u_2}{\partial a}\right|^2,\label{SUtwopot}\end{equation}
where
\begin{equation}W_{\text{eff}}(u_2)=g\left(\beta u_6(u_2)+\gamma u_4(u_2)+u_2\right).\end{equation}
Expressing the various ingredients of \eqref{SUtwopot} in terms of
$\tau$ is quite straightforward.  We have already seen that
\begin{equation}u_2=3b_2^2\wp(\tau/2)\end{equation}
and recall from \eqref{ringrel} that 
\begin{equation}u_4=\frac{u_2^2}{2}+\frac{\Lambda^4}{4},\qquad u_6=\frac{u_2^3}{3}+\frac{u_2\Lambda^4}{2}.\end{equation}
To determine $\partial u_2/\partial a$ will require a bit more work.  We
first compute the Seiberg-Witten 1-form in our parametric formalism
\begin{equation}d\lambda_{SW}=v\,ds=\left[-2b_2\left(F_1^{(2)}+F_2^{(2)}\right)+2b_2\left(\wp(\tau/2)-4\eta_1\right)\right]\,dz\end{equation}
in order to evaluate
\begin{equation}a=\oint_{\hat{A}_1}\,d\lambda_{SW}=2b_2\left(\wp(\tau/2)-4\eta_1\right).\end{equation}
We then compute
\begin{equation}\frac{\partial u_2}{\partial\tau}=\frac{b_2^2}{i\pi}\left(3\wp(\tau/2)^2-g_2\right),\qquad \frac{\partial a}{\partial\tau}=\frac{b_2}{2\pi i}\left(3\wp(\tau/2)^2-g_2\right)\end{equation}
from which it follows that
\begin{equation}\frac{\partial u_2}{\partial a}=2b_2.\end{equation}
Putting it all together, we find in the end that
\begin{equation}\frac{\partial u_2}{\partial a}\partial_{u_2}W(u_2)=gb_2\left(2+6\gamma b_2^2\wp(\tau/2)+\beta b_2^4\left(30\wp(\tau/2)^2-g_2\right)\right),\end{equation}
which we recognize as nothing other than twice the value of $A_2$ \eqref{Asbdrycond} required by our boundary conditions!  This means that $V_{SW}$ can simply be written as
\begin{equation}V_{SW}\sim \tau_2^{-1}|A_2|^2.\label{SWpottau}\end{equation}
A direct evaluation of the regulated integral $\int^{\Lambda_0}_{\Sigma}\,|dw|^2$ on the curve \eqref{OOPPsol} can easily be seen to yield an identical result, as expected from our general arguments in section \ref{sec:SW}.  Differentiating this with respect to $\tau$, we find that critical points occur whenever the following condition is satisfied
\begin{equation}2+6\gamma b_2^2\wp(\tau/2)+\beta b_2^4\left(30\wp(\tau/2)^2-g_2\right)=\frac{4\tau_2\left(3\wp(\tau/2)^2-g_2\right)\left(b_2^2\gamma + 6\beta b_2^4\wp(\tau/2)\right)}{2\pi + \tau_2\left(\wp(\tau/2)-4\eta_1\right)}.\end{equation}
This is precisely equivalent to the condition \eqref{compA2eqn} imposed by requiring that the curve \eqref{OOPPsol} arises via the OOPP limit of an exact solution of the form \eqref{SUtwoExacta}, \eqref{SUtwoExactb}.

\section{Mechanism for Metastability}
\label{sec:mechanism}

In this section, by taking a semiclassical limit of the exact curve
obtained in the previous section, we discuss the basic mechanism for the
metastability of the OOPP vacua.  We will see that the metastability is
the result of the balance between two forces, both of which can be
understood in a geometric manner.

\subsection{A Semiclassical Mechanics}

Let us consider the situation where the distance between the two tubes
corresponding to the two D4-branes is much larger than the size of the
tubes.  In this limit the Riemann surface of the M5-brane becomes a long
torus with $\tau_2\gg 1$.  If $\tau_2\gg 1$, using the $q$-expansion
formulas \eqref{qexpn} in \eqref{u2b2} and \eqref{b2Lambda}, we obtain
\begin{align}
 q=e^{i\pi \tau}={\Lambda^4\over 64u_2^2}.
\end{align}
Therefore, $\tau_2\gg 1$ corresponds to $u_2\gg \Lambda^2$, namely the
semiclassical limit of gauge theory.

In this semiclassical limit, the size of the tubes are
negligible and we can think of the system approximately as made of two
NS5-branes with D4-branes stretching between them.  The NS5-branes are
curved in the $w$ direction along $w=\pm \partial_v W(v)$ and, at the
same time, logarithmically bent in the $x^6$ direction so that the
distance $L$ between two NS5-branes is a function of $v$; see Figure
\ref{fig:semiclass}.
\begin{figure}[htbp]
 \begin{center}
  \epsfxsize=6cm \epsfbox{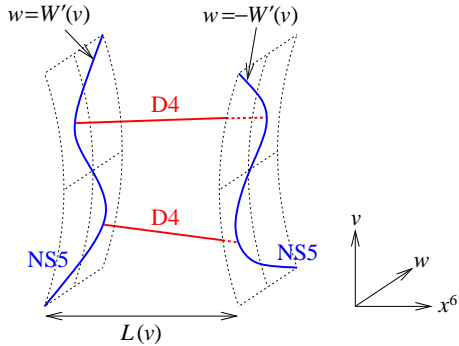}
 \end{center}
 \caption{A semiclassical picture of the configuration. The two NS5's
 are curved in the $w$ direction along $w=\pm W'(v)$ and bent in the
 $x^6$ directions such that the distance between them is $L(v)$.  We
 depicted this situation by the two NS5's being curved along curves
 $w=\pm W'(v)$ (blue) in two bent $w$-$v$ planes (dotted lines) which
 are at $x^6=\pm L(v)/2$.} \label{fig:semiclass}
\end{figure}
The relation between $s$ and $v$ is, from eq.\ \eqref{SWcurvess},
\begin{align}
 s&=\log\left({P_2(v)\pm\sqrt{P_2(v)^2-\Lambda^{4}}\over\Lambda^2}\right).
\end{align}
For $|v|\gg |\Lambda|$, this gives
\begin{align}
 s&\sim \begin{cases}
	 2\log (v/\Lambda),\\
	 2 \log(\Lambda/v).
	\end{cases}
\end{align}
Therefore, the distance along the $x^6=\Re s$ direction is given by
\begin{align}
 L&= \Delta x^6
 =R_{10} \Delta\Re s
 = 2R_{10}\log\left|{v\over\Lambda}\right|-2R_{10} \log\left|{\Lambda\over v}\right|
 = 2R_{10} \log\left|{v\over\Lambda}\right|^2.\label{L_semiclass}
\end{align}

Let us place one D4-brane at $v$ and the other at $-v$. Then the
D4-branes tilt in the $w$ direction by a small angle
\begin{align}
 \theta&\approx {2|\partial_v W|\over L},
\end{align}
as one can see from Figure \ref{fig:forcebalance}.\footnote{One may
wonder that the tilted D4-branes pull the NS5-branes also in the $w$
direction, so that the NS5-branes will not lie on the curve $w=W'(v)$
but on some ``distorted'' curve.  However, such an effect is of higher
order in $\theta$ and ignorable in the present approximation.
\label{ft:no_w-distorting}}
\begin{floatingfigure}{0.5\textwidth}
\begin{center}
\epsfig{file=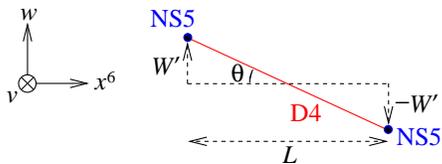,width=0.35\textwidth}
\caption{D4-branes tilted along $w$}
\label{fig:forcebalance}
\end{center}
\end{floatingfigure}
The endpoints of D4-branes on the same NS5-brane interact with each
other through a Coulomb interaction in the NS5 worldvolume.  Because a
D4-brane endpoint has codimension 2 in an NS5-brane worldvolume, the
Coulomb potential goes logarithmically with the distance.  In the
present case, the D4-brane at $v$ and the one at $-v$ have a Coulomb
potential energy $V_{C}$.  In addition, there is potential energy coming
from the tension of the D4-branes, which is $V_{T}=2T_4 L$.  Here, $T_4$
is the tension of a D4-brane and the factor $2$ is because we have two
D4-branes.
If the D4-branes were not tilted in the $w$ direction, {\it i.e.}, if
$\theta=0$, then the system would be supersymmetric and the two
potentials would exactly cancel each other: $V_C+V_T^{\theta=0}=0$.
However, if $\theta\neq 0$, the D4-branes are longer by $\Delta
L=(L/\cos\theta)-L\approx L\theta^2 /2$ than they are for $\theta=0$.
As the result, $V_T$ gets increased by
\begin{align}
 \Delta V_T=V_T^{\theta}-V_T^{\theta=0}=2T_4\, \Delta L\approx  {T_4\theta^2\over L}
 \approx {4T_4|\partial_v W|^2\over L}.
\end{align}
On the other hand, $V_C$ is not affected by $\theta$ because the Coulomb
force depends only on the distance between D4-brane endpoints.
Therefore, the total potential energy when $\theta\neq 0$ is given by
\begin{align}
 V_{sc}=V_C+V_T^{\theta}={4T_4|\partial_v W|^2\over L}.
 \label{semiclass_pot}
\end{align}
The equation of motion derived from this potential energy is
\begin{align}
 0=\p_v V_{sc}
 \propto {\p_v^2 W\over \p_v W}-{\p_v L\over L}.
\label{semiclass_eom}
\end{align}
One can interpret the first term in \eqref{semiclass_eom} as a force
which pushes the D4-branes towards the values of $v$ for which $|\p_v
W|$ is smaller, and the second term as a force towards the values of $v$
for which $L$ is larger.  As can be seen from \eqref{L_semiclass}, the
latter force is logarithmic and tend to move the D4-branes toward large
$|v|$.  On the other hand, the former force is polynomial and can be
tuned by choosing the polynomial $W(v)$.  Therefore, it is natural to
expect that, for any value of $v$, one can choose the superpotential
$W(v)$ appropriately so that the equation of motion
\eqref{semiclass_eom} is satisfied for that $v$.  This is the geometric
understanding of the metastable vacua found in \cite{Ooguri:2007iu}.

As a quick check, let us compare the potential \eqref{semiclass_pot}
obtained above with the potential energy in gauge theory.  In the
semiclassical regime where $|u_2|\gg|\Lambda|^2$, the moduli space
metric for $u_2$ is known to be \cite{Seiberg:1994rs}:
\begin{align}
 g^{u_2\overline u_2}&\approx {2\pi |u_2|\over \log\left|{u_2\over\Lambda^2}\right|}.
\end{align}
Therefore, the gauge theory potential is:
\begin{align}
 V_{gt}&=g^{u_2\bar u_2}|\partial_{u_2}W|^2
 \approx{2\pi |u_2|\over \log\left|{u_2\over \Lambda^2}\right|}|\partial_{u_2}W|^2.
\end{align}
If we rewrite this in terms of $v=\sqrt{u_2}$,
\begin{align}
 V_{gt}&={\pi \over 2\log\left|{v\over \Lambda}\right|^2}|\p_v W|^2,
\end{align}
which agrees with the semiclassical potential \eqref{semiclass_pot} up
to a numerical factor if we use \eqref{L_semiclass}.  So, the potential
\eqref{semiclass_pot} is indeed correct in the semiclassical regime.

One can also understand the parameter $\nu$ that appears in the exact M5
curve by a semiclassical reasoning, although strictly speaking $\nu$ is
vanishing in the OOPP limit ($\nu=\CO(g^2)$).  If a D4-brane parallel to
the $x^6$-axis is ending on an NS5-brane, then the lift, an M5-brane,
has the following curve:
\begin{align}
 x^6&= 2R_{10}\log|v|=R_{10}(\log v+\log\vb),\qquad
 x^{10}= 2R_{10}\arg|v|=-i R_{10}(\log v-\log\vb).\label{x6x10}
\end{align}
In the NS5 language, the $v$-dependence of $x^6$ represents the pull of
the tension of the D4-brane along the $x^6$ direction, while the
$v$-dependence of $x^{10}$ represents the flux inserted into the NS5
worldvolume by the D4-brane.
Instead, if the D4-brane is not parallel to the $x^6$-axis but makes a
small angle $\theta$ with it, then the tension will reduce by factor
$\cos\theta$ and hence \eqref{x6x10} is replaced by
\begin{align}
 x^6&= R_{10}\cos\theta(\log v+\log\vb),
\end{align}
whereas $x^{10}$ is unchanged because the amount of flux inserted by a
D4 is independent of the angle between the D4 and the NS5.  Therefore,
$s=x^6+ix^{10}$ goes as
\begin{align}
 s&= R_{10}[(\cos\theta+1)\log v+(\cos\theta-1)\log\vb]
 \approx R_{10}\left[\left(2-{\theta^2\over 2}\right)\log v-{\theta^2\over 2}\log\vb\right].
\end{align}
Comparing this with the expression of the exact curve \eqref{SUtwoExacta}, we
obtain
\begin{align}
 \nu&= -{\theta^2\over 2}=-{2|\p_v W|^2\over L^2}.\label{nu_semiclass}
\end{align}
We will see below that this is indeed satisfied in our M5-brane curve.

\subsection{Semiclassical Limit of the M5 Curve}

In the previous subsection, we showed that, in general, the
semiclassical potential $V_{sc}$ (eq.\ \eqref{semiclass_pot}) and the
gauge theory potential $V_{gt}$ are the same for $|v|\gg |\Lambda|$.
Here we will check that the M5 curve obtained in section
\ref{sec:curves} satisfies the equations of motion derived from these
potentials.

In the M5 curve \eqref{OOPPsol}, the quantities $\tau,\nu$ are
determined in terms of the parameters $A_6$, $A_4$, $A_2$, $b_2$, or
equivalently in terms of $(g,\beta,\gamma)$, $b_2$, by solving the last
two equations in \eqref{SUtwoExacta}.  In the semiclassical limit
$\tau_2\gg 1$, $q=e^{i\pi \tau}$ is very small and we can use the
$q$-expansions \eqref{qexpn} to simplify those equations.  After
dropping all $\CO(q)$ terms and subleading terms in $g$, we can write
the resulting equations as follows:
\begin{subequations}
\begin{align}
  \tau_2
 &={2(\beta v^4+\gamma v^2+1)\over \pi(5\beta v^5+3\gamma v^2+1)}
 ={2\over \pi}{\p_v W(v)\over v\, \p_v^2 W(v)}
 \label{tau2scM}
 \\
 \nu
 &=-{|gv|^2\over 2R_{10}^2}{(\overline{5\beta v^5+3\gamma v^2+1})^2(\beta v^4+\gamma v^2+1)\over \overline{\beta v^4+\gamma v^2+1}}
 =-2\left({|\p_v W(v)|\over \pi g_2 \tau_2}\right)^2,
 \label{nuscM}
\end{align}
\end{subequations}
where we have set $v=i\pi b_2$ and $\p_v W=g(\beta v^5+\gamma v^3+v)$.
Semiclassically, D4-branes are sitting at $v$ and $-v$.  

Also, using the $q$-expansion \eqref{qexpn} in \eqref{b2Lambda}, we
obtain $q=\Lambda^4/64\pi^4b_2^4$.  Therefore, we can express $\tau_2$
in terms of $v$ as follows:
\begin{align}
 \tau_2
 ={1\over \pi}\log\left|{64\pi^4 b_2^4\over\Lambda^4}\right|
 ={1\over \pi}\log\left|{64v^4\over\Lambda^4}\right|
 \approx {2\over \pi}\log\left|{v\over\Lambda}\right|^2.
\end{align}
Therefore, \eqref{tau2scM} can be written as
\begin{align}
 {\p_v^2 W(v)\over \p_v W(v)}&={2\over \pi v\tau_2}
 \approx{1/v\over \log\left|{v\over\Lambda}\right|^2}.
\end{align}
If we recall the expression for $L$, eq.\ \eqref{L_semiclass}, we
immediately see that this is nothing but \eqref{semiclass_eom}.  So,
indeed, the M-theory curve satisfies the semiclassical equation of
motion derived from $V_{sc}$ or $V_{gt}$.

Similarly, one can show that the value of $\nu$ in the M5 curve given by
\eqref{nuscM} agrees with the one given by the semiclassical expression
\eqref{nu_semiclass}.\footnote{For this agreement, the numerical
coefficient in \eqref{nu_semiclass} is important.  As mentioned in
footnote \ref{ft:no_w-distorting}, \eqref{nu_semiclass} was derived
assuming that there is no ``distortion'' of the NS5-branes in the $w$
direction.  Therefore, this agreement means that the approximation to
ignore the ``distortion'' was consistent.}
These agreements confirm that the reason for the existence of
(meta)stable M5-brane configurations is the geometric mechanism discussed
in the previous subsection.

As we show in Appendix \ref{app:IIAlim}, we can also take the type IIA
limit of the M5 curve, which is defined as the $R_{10}\to 0$ limit with the
distance between the NS5's kept fixed.  In this limit, we end up with
two NS5's with D4's stretching between them, and furthermore the log
bending of NS5's in the $x^6$ direction disappears.
In order for the distance between the two D4's to remain finite in the
limit, we must take $\tau_2\to\infty$ as $\tau_2\sim 1/R_{10}$, as one can
see in \eqref{tau2scaleasgs}.  Therefore, \eqref{tau2scM} in this limit
means that the D4's sit at $v$ for which $\partial_v^2 W(v)=w'(v)=0$.
This is understood as follows: In the IIA limit, there is no log bending
of NS5's along $x^6$ and the associated force, the second term in
\eqref{semiclass_eom}, is not there.  So, the D4's must sit at points
where they can sit in equilibrium without any force (although unstable
in this limit), which happens for $w'(v)=0$.  If one makes $R_{10}\neq 0$
and the quantum effects of bending are turned on, though, these points
can be stabilized if the superpotential is properly tuned.

\section{Concluding Remarks}\label{sec:conclusion}

In this paper, we demonstrated that the M5 lifts of NS5/D4
configurations which realize perturbed Seiberg-Witten theory accurately
capture, in a suitable limit, the full scalar potential of the gauge
theory and hence all nonsupersymmetric vacua that it describes.  Within
this regime, M5 configurations that approximately solve the equations of
motion can be found and take the simple form of a holomorphic
Seiberg-Witten geometry harmonically embedded along a transverse
direction.  These represent stringy realizations of the metastable vacua
described by \cite{Ooguri:2007iu,Pastras:2007qr}.  Crucial to this
story, however, was the fact that the scalar potential emerged only
after the M5 worldvolume action was properly regularized.  As such, the
approximate solutions are not valid all the way out to $\infty$, but
rather only within an appropriately chosen cutoff surface.  We were able
to make this completely explicit for the simple case of $SU(2)$ gauge
group by finding exact solutions and demonstrating the existence of the
appropriate ``OOPP limit.''

Our $SU(2)$ example also allowed us to obtain a geometric understanding
for the gauge theory potential, at least in the semiclassical limit, as
well as the mechanism for stability.  Stated in IIA language, the
logarithmic bending of NS5's by the D4's which end on them creates
pockets where the D4's want to remain in order to minimize their length.
The tendency to sit in these pockets, however, receives competition from
the logarithmic repulsion of the D4's stemming from their interaction
through the NS5 worldvolumes.  Nevertheless, one can suitably tune the
asymptotic NS5 geometry in order to achieve stability.  This is
precisely the OOPP mechanism \cite{Ooguri:2007iu, Pastras:2007qr} at
work.

It should be noted that the story described here is quite similar to
that considered in \cite{ourpaper}, where a different limit was studied.
In that regime, the bending of the NS5's is quite large and the lightest
modes are fluctuations of the tubes into which the D4's blow up in the
M5 lift.  The asymptotic $w(v)$ geometry is thus considered to be rigid
with only $s(v)$ dynamical{\footnote{This is equivalent to what we
called the Dijkgraaf-Vafa regime earlier in section \ref{sec:DV}, though
the configurations of \cite{ourpaper} contained both D4's and
\anti{D4}'s so are not relevant for describing a gauge theory.}}.
Applying a $T$-duality along $x^6$ to this system, one obtains a
deformed Calabi-Yau geometry determined by $w(v)$ with R and NS field
strengths dual to $s(v)$.  The limit of \cite{ourpaper}, then,
corresponds on the IIB side to one in which the presence of nontrivial R
and NS fluxes generates a potential for the moduli of a rigid
Calabi-Yau.  This potential can be computed in IIB
\cite{Aganagic:2006ex} and is precisely equivalent to that which follows
from the M5 worldvolume action along the lines of \cite{ourpaper} and
section \ref{sec:DV}.  In light of this, it is natural to ask whether
the scalar potential of perturbed Seiberg-Witten theory studied in this
paper also has a type IIB counterpart.  The most naive approach, namely
$T$-dualizing along $x^6$, leads to an awkward regime in which
nonnormalizable deformations of the geometry play a role in generating
the potential for complex structure moduli rather than the fluxes.
Applying $T$-duality along other directions, though, can lead to more
natural constructions that are currently under investigation
\cite{futurework}.

\section*{Acknowledgments}

We would like to thank J.~de~Boer, R.~Dijkgraaf, L.~Mazzucato,
H.~Ooguri, C.~S.~Park, K.~Skenderis, M.~Taylor and especially
Y.~Ookouchi for valuable discussions.  We would also like to thank
Y.~Ookouchi for collaboration at the early stage of this work.
The work of J.M. was supported in part by Department of Energy grant
DE-FG03-92ER40701 and by a John A McCone postdoctoral fellowship.  The
work of K.P. was supported by Foundation of Fundamental Research on
Matter (FOM).  The work of M.S. was supported by an NWO Spinoza grant.

\appendix

\section{Some Basic Results on Riemann Surfaces}\label{app:RiemannS}

In this appendix we summarize a few basic properties of Riemann surfaces, for
more details see \cite{grifhar}. 

A compact Riemann surface $\Sigma_g$ is a one-dimensional 
compact complex manifold.
Its topology is completely characterized  by an integer, the genus $g$. 
The middle cohomology group has
dimensionality $\dim H^1(\Sigma_g) = 2g$. The intersection form on 
$H_1(\Sigma_g,\mathbb{Z})$ is antisymmetric and by Poincar\'e duality unimodular, 
which means that we can  pick  a basis of one-cycles $A_i,B_j$ with intersection:
\begin{equation}
  A_i\cap A_j =0,\quad B_i\cap B_j = 0,\quad A_i \cap B_j = \delta_{ij}.
\end{equation}
Such a basis is unique up to a symplectic transformation in $Sp(2g,\mathbb{Z})$,

A surface $\Sigma_g$ of genus $g$ has a complex structure moduli space 
${\cal M}_g$ of dimensionality $\dim{\cal M}_g = 3g-3, \, g\geq 2$. 

A 1-form $\omega$ on a Riemann surface is called a 
\emph{holomorphic differential} if in a local coordinate patch it has the form:
\begin{equation}
  \omega = f(z) dz
\end{equation}
with $f(z)$ holomorphic. We will also consider \emph{meromorphic differentials},
for which we allow the function $f(z)$ to have poles at certain 
points on the surface. Now we present a standard basis for holomorphic and
meromorphic differentials on a general Riemann surface:

\paragraph{Holomorphic differentials\protect\footnote{These are also called
meromorphic differentials of the first kind.} $\omega_i$:} Once we pick
a symplectic basis of one-cycles, there is a canonical basis of
holomorphic differentials $\omega_i,i=1,..,g$, with the following
periods:
\begin{equation}
  \oint_{A_i} \omega_j = \delta_{ij},\qquad \oint_{B_i} \omega_j = \hat{\tau}_{ij}.
\end{equation}
The (symmetric) matrix $\hat{\tau}_{ij}$ is the period matrix of the surface,
and depends on the complex structure of $\Sigma_g$.

\paragraph{Meromorphic differentials of the second kind, $d \Omega_{n\ge 1}^P$:}
These are characterized by a point $P$ on the surface where the 
differential has a pole of order $n+1$ with $n\geq 1$. The are normalized
so that in local complex coordinates $z$ where $z(P) = 0$ they have 
the Laurent expansion:
\begin{equation}
  d\Omega^P_n \sim  n{dz \over z^{n+1}}  + \text{regular}.
\label{meroexpand}
\end{equation}

\paragraph{Meromorphic differentials of the third kind, $d \Omega_0^{P,P'}$:} 
characterized by two points $P,P'$, where the differential has first 
order poles with opposite residues. Around $P$ we have:
\begin{equation}
  d\Omega_0^{P,P'} \sim {dz\over z}  + \text{regular}
\label{meroexpandb}
\end{equation}
and similarly around $P'$ with the opposite sign.

Notice that we can always shift a meromorphic differential by
a holomorphic differential without changing the singular part of the 
Laurent expansions \eqref{meroexpand}, \eqref{meroexpandb}. We can eliminate 
this ambiguity by demanding that the $A$ periods of the meromorphic
differentials vanish:
\begin{equation}
  \oint_{A_i} d\Omega_n^P = 0.
\end{equation}
In general, it is not possible to simultaneously set the $B$ periods 
to zero. Instead we have:
\begin{equation}
  \oint_{B_i} d\Omega_n^P = K^P_{in},
\end{equation}
where the matrix $K^P_{in}$ depends on the complex structure moduli of the
Riemann surface and the position of the puncture $P$.

\paragraph{Harmonic functions on Riemann surfaces:} In two dimensions a function
is harmonic if it satisfies $\partial\overline{\partial}
f(z,\overline{z}) = 0$.  Locally it is a sum of a holomorphic and an
antiholomorphic function.  On a compact surface, a harmonic function is
necessarily constant. If we allow for poles at points $\{ P_a\}$ we can
also have nonconstant harmonic functions.  They can be constructed using
the meromorphic differentials. Consider the 1-form:
\begin{equation}
  \omega = \sum_{a}\sum_{n\geq 1} T_n^{a} d \Omega^{P_a}_n + \sum_{a,b}T_0^{ab}
d\Omega^{P_a, P_b}_0+ \sum_i h_i \omega_i + \text{c.c.}
\label{genharmonf}
\end{equation}
Given $\omega$, locally we can always find a function $f$ such that $\omega=df$,
and it is easy to see that $f$ is harmonic.  To make sure the $f$ can be
globally well defined on the Riemann surface, we must make sure that the
compact periods of $\omega$ vanish:
\begin{equation}
  \oint_{A_i}\omega= 0,\quad \oint_{B_i}\omega = 0
\end{equation}
and that the total residue at each of the points $P_a$ vanishes.

\section{Parametric Representation of Genus 1 Seiberg-Witten Geometry}\label{app:SWparam}

In this appendix, we describe the parametric representation of the genus
1 M5 curve of section \ref{subsec:SWparam} in more detail as a means of
demonstrating basic techniques for manipulating the elliptic functions
$F_i^{(n)}$\,{\footnote{See \cite{ourpaper} for a detailed description of
the $F_i^{(n)}$ and their properties.}}.  In particular, we focus on the
M5 lift of an NS5/D4 configuration of the type depicted in figure \ref{SWNS5D4} with two
parallel NS5's extended along $v$ and two D4-branes suspended in between.
The M5 lift has genus 1 and can be described by two embedding functions
$s(z)$ and $v(z)$ defined on the fundamental parallelogram of figure
\ref{zplane}.  As discussed in section \ref{subsec:SWparam}, $s(z)$ and
$v(z)$ take the form
\begin{equation}\begin{split}v(z)&=b_2\left(F_1^{(1)}-F_2^{(1)}+i\pi\right),\\
s(z)&=2\left(F_1-F_2+i\pi z\right).
\label{vsembeds}\end{split}\end{equation} Moreover, one of the
$\hat{B}$-period constraints for $ds$ fixes the distance between the
marked points in terms of $\tau$
\begin{equation}a\equiv a_2-a_1=\frac{\tau}{2}.\end{equation}
The other $\hat{B}$-period constraint determines how the dynamical scale
$\Lambda$ is related to curve parameters.  Rather than studying the
$\hat{B}$-period constraint directly, though, let us instead try to
determine the explicit relationship between $v(z)$ and $s(z)$ in
\eqref{vsembeds} and read it off from the Seiberg-Witten geometry
(recall that $t\equiv \Lambda^N e^{-s}$)
\begin{equation}t^2-2P_N(v)t+\Lambda^{2N}=0.\end{equation}

It is easiest to work with elliptic functions so, to start, we study the
derivative of $s$.  This leads us to observe that
\begin{equation}
ds =
2\left(F_1^{(1)}-F_2^{(1)}+i\pi\right)dz=\frac{2v\,dz}{b_2}.\label{dsvexp}\end{equation}
From this, it is clear that we can easily determine $s(v)$ by
integration once we are able to write $\frac{dv}{dz}$ as a function of
$v$.  As such, we turn to
\begin{equation}G(z)\equiv \frac{dv}{dz} = b_2\left(F_1^{(2)}-F_2^{(2)}\right)\end{equation}
and seek an algebraic relationship between the elliptic functions $G(z)$
and $v(z)$.  Because $G(z)$ has second order poles while $v(z)$ has
first order ones, our first guess might be that $G(z)$ is given by a
quadratic polynomial in $v(z)$.  This is impossible, though, because the
second order poles of $G(z)$ have opposite signs while those of $v(z)^2$
have the same sign.  Because of this, we instead consider the
possibility that $G(z)^2$ is equivalent to a quartic polynomial in
$v(z)$\,{\footnote{The overall factor of $b_2^2$ can already be seen from
the definitions of $v(z)$ and $G(z)$.  The symmetry under
$v\leftrightarrow -v$ can also be verified ahead of time along the lines
of appendix \ref{app:symm}.}}
\begin{equation}G(z)^2-b_2^2\left(v^4+C_1v^2+C_2\right)=0.\end{equation}
Because the LHS of this equation is an elliptic function, it is
completely specified, up to a constant shift, by its pole structure.  As
such, we need only verify that it vanishes at the marked points $a_1$
and $a_2$.  The coefficients of the poles at $a_1$ and $a_2$ are
equivalent up to possible minus signs, though, so we only need to study
$G(z)^2$ and $v(z)^m$ in the vicinity of one marked point, say $a_1$.
Expanding $G(z)^2$, we find
\begin{equation}\begin{split}G(z)^2&= b_2^2\left(F_1^{(2)}-F_2^{(2)}\right)^2\\
&\sim\frac{b_2^2}{(z-a_1)^4}-\frac{2b_2^2\wp(\tau/2)}{(z-a_1)^2}+b_2^2\left(\frac{3g_2}{5}-5\wp(\tau/2)^2\right)+{\cal{O}}(z-a_1).\end{split}\end{equation}
On the other hand,
\begin{multline}
 \frac{1}{b_2^2}\left(v^4+C_1v^2+C_2\right)
 \sim
 \frac{b_2^2}{(z-a_1)^4}
 +\frac{C_1+4b_2^2\wp(\tau/2)}{(z-a_1)^2}\\
 +\left(\frac{C_2}{b_2}^2-\frac{2b_2^2g_2}{5}+2C_1\wp(\tau/2)+10b_2^2\wp(\tau/2)^2\right)
 +{\cal{O}}(z-a_1).
\end{multline}
From this, we see that
\begin{equation}G(z)^2=\frac{1}{b_2^2}\left[\left(v(z)^2-3b_2^2\wp(\tau/2)\right)^2+b_2^4\left(g_2-12\wp(\tau/2)^2\right)\right].\end{equation}
Returning to \eqref{dsvexp}, this implies that{\footnote{The branch of the square root that we use when writing $G(z)$ is correlated with how we choose to identify the marked points $a_1$ and $a_2$ with the points at $\infty$ on the two sheets covering the $v$-plane.  By convention, we take $a_1$ to correspond to the point at $\infty$ on the first sheet, hence the minus sign in \eqref{dsdv}.}}  
\begin{equation}ds=-\frac{2v\,dv}{\sqrt{\left(v^2-3b_2^2\wp(\tau/2)\right)^2+b_2^4\left(g_2-12\wp(\tau/2)^2\right)}},\label{dsdv}\end{equation}
which can be integrated to yield
\begin{equation}s(v)=-\ln\left(v^2-u-\sqrt{(v^2-u)^2-\Lambda^4}\right)+\text{constant},\label{svres}\end{equation}
where
\begin{equation}u=3b_2^2\wp(\tau/2)\qquad \Lambda^4=b_2^4\left(12\wp(\tau/2)^2-g_2\right).\label{uLids}\end{equation}
Our choice of notation $u$ and $\Lambda$ is already quite suggestive.
Indeed, it is easy to demonstrate now that, dropping the constant in
\eqref{svres}, $t=\Lambda^2 e^{-s}$ satisfies
\begin{equation}t^2-2P_2(v)t+\Lambda^4=0\end{equation}
for
\begin{equation}P_2(v)=v^2-u.\end{equation}
This justifies our identification of $\Lambda$ in \eqref{uLids}.  In
particular, we see that the constant $b_2$ is now fixed in terms of
$\tau$ and $\Lambda$.  The only free parameter left in our solution,
then, is $\tau$.  While this provides a perfectly fine parametrization
of the moduli space, a more conventional one is given by $u_2$
\eqref{updef}.  The relation between $u_2$ and $\tau$ can be read off
directly from \eqref{uLids} as $u_2$ should be identified the parameter
$u$ there.  We can also use \eqref{upcomp}, however, to compute
$u_2(\tau)$ directly from our parametric description \eqref{vsembeds}
\begin{equation}\begin{split}u_2&=\frac{1}{4\pi i}\oint_{a_1}v^2\,ds\\
&=\frac{1}{4\pi i}\oint_{a_1}\,2b_2^2\left(F_1^{(1)}-F_2^{(1)}+i\pi\right)^3\,dz.\end{split}\end{equation}
This can be easily evaluated by expanding $F_1^{(1)}$ and $F_2^{(1)}$ near $a_1$ and results in
\begin{equation}u_2=3b_2^2\wp(\tau/2).\end{equation}
Consequently, $u_2$ is nothing other than the parameter $u$ in \eqref{uLids}.

\section{Symmetry Argument for the Ansatz \eqref{SUtwoExacta}}
\label{app:symm}

In this Appendix, we motivate the ansatz \eqref{SUtwoExacta} for the M5
curve describing the OOPP vacua from a symmetry argument.

First, let us examine the symmetry of the M5 curve in the supersymmetric
case.  As explained in section \ref{sec:IIASW}, an M5 curve
corresponding to supersymmetric vacuum of $SU(2)$ theory has the
following form:
\begin{gather}
 t^2-2P_N(v)t+\Lambda^{2N}=0,\label{susycrvt}\\
 w(v)=\sqrt{W_n'(v)^2-f_{n-1}(v)}\label{susycrvw},
\end{gather}
where $t=\Lambda^{N}e^{-s}$, and satisfies the constraint
\eqref{facteqn} or \eqref{facteqnn}. Because in the $SU(2)$ case
$\Tr\Phi^k=0$ for odd $k$, $P_2(v)=v^2-u_2$ and $W_n'(v)$ has only odd
powers in $v$.  By examining \eqref{facteqn}, \eqref{facteqnn}, one can
see that this implies that $f_{n-1}(v)$ has only even powers in $v$.
Accordingly, under $v\to -v$,
\begin{align}
 P_2(v)&\to P_2(v),\qquad
 W'_n(v)\to -W'_n(v),\qquad
 f_{n-1}(v)\to f_{n-1}(v).
 \label{symPWf}
\end{align}
From \eqref{susycrvt}, one obtains
\begin{align}
 s(v)=\ln\left({P_2(v)+\sqrt{P_2(v)^2-\Lambda^{4}}\over \Lambda^2}\right).\label{susycrvs}
\end{align}
The relations \eqref{susycrvw} and \eqref{susycrvs} define the $w,s$
coordinates as functions on the two-sheeted cover of the complex $v$ plane.

This supersymmetric curve has the following symmetries:
\begin{equation}
 \begin{split}
 \text{symmetry I)}  &\quad  v\to -v \text{~(same sheet)},\quad      w\to -w, \quad s\to s;\\
 \text{symmetry II)} &\quad  v\to  v \text{~(different sheet)},\quad w\to -w, \quad s\to -s.
\end{split}
\label{susysym}
\end{equation}
For example, if we flip $v\to -v$ remaining on the same sheet, then from
\eqref{susycrvw}--\eqref{susycrvs} it is easy to see
that $w=\sqrt{W_{n}'{}^2+f_{n-1}}=W_{n}'+{f_{n-1}\over 2W_{n}}+\cdots\to
-W_{n}'-{f_{n-1}\over 2W_{n}}-\cdots=-w$ and $s\to s$ (symmetry I).  On
the other hand, if move $v$ between the first and second sheets, the
square roots in \eqref{susycrvw} and \eqref{susycrvs} flip their signs
and we have $w\to-w$, $s\to -s$ (symmetry II).

Because we will write M5 curves in parametric representation on the $z$
plane, we need to implement these symmetries \eqref{susysym} on the $z$
plane.  In the supersymmetric case, the relation between $v$ and $z$ is
given by \eqref{vSW}.  Using the properties of $F^{(1)}_i(z)$ in
\eqref{Fmonods}, one can see that, if $a=a_2-a_1=\tau/2$,
\begin{align}
\begin{array}{r@{\,}lcr@{\,}l}
 \text{transformation I) }\qquad v&\to -v\text{~(same sheet)}     &\Longleftrightarrow & z &\to 2a_1-z \cong 2a_2-z,\\[.5ex]
 \text{transformation II)}\qquad v&\to v\text{~(different sheet)} &\Longleftrightarrow & z &\to a_1+a_2-z.
\end{array}
\label{ztrfm}
\end{align}

Let us require that the exact M5 curve corresponding to the OOPP vacuum
have the same symmetries as above. Namely, under the transformations of
$z$ in \eqref{ztrfm}, we require that $s,v,w$ transform according to
\eqref{susysym}.  The 1-forms $ds,dv,dw$ have the same transformation
property as $s,v,w$.
The basis of 1-forms $\omega$, $d\Omega_0^{a_1,a_2}$,
$d\Omega_{n\ge 1}^{a_i}$, $i=1,2$ introduced in section
\ref{subsec:harmfuncsT2} have the following transformation property
under the transformations \eqref{ztrfm}:
\begin{equation}
\begin{split}
  \text{I)} \quad
 & \omega\to -\omega,\quad
 d\Omega_0^{a_1,a_2}\to +d\Omega_0^{a_1,a_2},\quad
 d\Omega_n^{a_1} \to (-1)^n d\Omega_n^{a_1},\quad
 d\Omega_n^{a_2} \to (-1)^n d\Omega_n^{a_2},\\
 \text{II)} \quad
 & \omega\to -\omega,\quad
 d\Omega_0^{a_1,a_2}\to -d\Omega_0^{a_1,a_2},\quad
 d\Omega_n^{a_1} \to (-1)^n d\Omega_n^{a_2},\quad
 d\Omega_n^{a_2} \to (-1)^n d\Omega_n^{a_1}.
\end{split}
\end{equation}
Therefore, in order to obey \eqref{susysym}, the 1-forms $ds,dv,dw$ must
have the following schematic form:
\begin{equation}
\begin{split}
  dw&\sim \sum_{n=1,3,5,\dots} (d\Omega_n^{a_1}+d\Omega_n^{a_2})+\omega+{\rm c.c.},\\
  dv&\sim \sum_{n=1,3,5,\dots} (d\Omega_n^{a_1}-d\Omega_n^{a_2})+{\rm c.c.},\\
  ds&\sim d\Omega_0^{a_1,a_2}+{\rm c.c.},
\end{split}
\end{equation}
where coefficients are omitted.  In other words,
\begin{equation}
\begin{split}
 dw&\sim \sum_{n=2,4,6,\dots} (F_1^{(n)}+F_2^{(n)})dz+dz+{\rm c.c.},\\
 dv&\sim \sum_{n=2,4,6,\dots} (F_1^{(n)}-F_2^{(n)})dz+{\rm c.c.},\\
 ds&\sim (F_1^{(1)}-F_2^{(1)}+i\pi)dz+{\rm c.c.}
\end{split}
\end{equation}
So far we have not taken into account any period constraints but, in
$dw$, $(F_1^{(2)}+F_2^{(2)})dz$ must come with $dz-d\zb$ for the period
along $\hat B_2-\hat B_1$ to vanish.  Therefore, $dw$ must come with
\begin{align}
 (F_1^{(2)}+F_2^{(2)})dz+{4\pi i\over\tau-\taub}(dz-d\zb).
\end{align}

In the small superpotential case studied in section \ref{sec:SW}, $v(z)$
was given by \eqref{vSW} and $dv(z)$ had order two poles at $z=a_{1,2}$.
To consider the OOPP vacuum, we would like to take a degree six
superpotential, which leads to the boundary condition \eqref{holobdcd}.
This motivates us to study $dw(z)$ that has up to order six poles at
$z=a_{1,2}$ and therefore contains $F^{(n)}_i$ with $n\le 6$.
However, as was discussed below \eqref{virconn}, the pole structure of
$v(z)$ given by \eqref{vSW} is not enough; in order to satisfy the
Virasoro constraint we need to include higher order poles in $dv(z)$ than
was considered in \eqref{vSW}.  In the present case, order four poles
are sufficient.  
After all, we are led to the following ansatz:
\begin{equation}
\begin{split}
 dw&\sim (F_1^{(6)}+F_2^{(6)})dz+(F_1^{(4)}+F_2^{(4)})dz
 +(F_1^{(2)}+F_2^{(2)})dz+{4\pi i\over\tau-\taub}(dz-d\zb),\\
 dv&\sim (F_1^{(4)}-F_2^{(4)})dz+(F_1^{(2)}-F_2^{(2)})dz+{\rm c.c.},\\
 ds&\sim (F_1^{(1)}-F_2^{(1)}+i\pi)dz+{\rm c.c.},
\end{split}
\end{equation}
which upon integration becomes the ansatz \eqref{SUtwoExacta} that we used.

\section{$q$-expansions}\label{app:q-expn}

When $\tau_2\gg 1$, $q=e^{i\pi \tau}=e^{i\pi \tau_1-\pi \tau_2}$ is
small and the following $q$-expansions \cite{ourpaper, Lang} are useful:
\begin{equation}
 \begin{split}
  F(z)&=\ln(e^{2\pi i z}-1)+\ln\left[1+\sum_{k=1}^\infty (2k+1)(-1)^k q^{k(k+1)}\right]
  +4\sum_{k=1}^\infty {q^{2k}\sin^2(k\pi z)\over k(1-q^{2k})},\\
 \wp\left({\tau\over 2}\right)&=-\frac{\pi^2}{3}-8\pi^2\sum_{k=1}^{\infty}\frac{kq^{k}}{1+q^{k}},\qquad\qquad
 \eta_1=\frac{\pi^2}{6}-4\pi^2\sum_{k=1}^{\infty}\frac{kq^{2k}}{1-q^{2k}}, \\
 g_2&={4\pi^4\over 3}\left(1+240\sum_{k=1}^{\infty}\frac{k^3q^{2k}}{1-q^{2k}}\right),\qquad\qquad
 g_3={8\pi^6\over 27}\left(1-504\sum_{k=1}^{\infty}\frac{k^5 q^{2k}}{1-q^{2k}}\right).
\end{split}
\label{qexpn}
\end{equation}
Note also that $\eta_1,\eta_3$ are related by $\eta_1\tau-\eta_3=i\pi$.

\section{Type IIA Limit of the M5 Curve}
\label{app:IIAlim}

In this Appendix, we consider the type IIA limit of the M5 curve in the
OOPP limit obtained in section \ref{subsec:pertN=2}.  The IIA limit is
defined to be the $R_{10}\to 0$ limit with the distance between NS5's and
that between D4's kept finite.  We will see that, in the limit, we end
up with curved NS5-branes along $w=\pm W'(v)$ with D4-branes stretched
in between, and that there is no logarithmic bending of NS5-branes along
the $x^6$ direction.  As a result, we will see that D4-branes sit at the
critical point of $w(v)$, namely at $v$ for which $W''(v)=0$.

\bigskip
The M5 curve in the OOPP limit obtained in section \ref{subsec:pertN=2}
can be written as:
\begin{equation}
 \begin{split}
  s&=2\left[F_1-F_2+i\pi(z-A-a) \right],\\
  v &=b_2(F_1^{(1)}-F_2^{(1)}+i\pi),\\
  w&=A_6(F_1^{(5)}+F_2^{(5)})+A_4(F_1^{(3)}+F_2^{(3)})
  +A_2\left[(F_1^{(1)}+F_2^{(1)})
  +{4\pi i\over\tau_2}\Im(z-A-a)
 \right].
\end{split}\label{svw}
\end{equation}
where the parameters $A_{2,4,6}$ are related to the coefficients
$g,\beta,\gamma$ in the superpotential \eqref{singltr} and
\eqref{holobdcd} via \eqref{Asbdrycond}.  Given these parameters, the
modulus $\tau$ is determined by solving \eqref{compA2eqn}.  Recall also
that $a\equiv a_2-a_1={\tau/ 2}$ and $A\equiv (a_1+a_2)/2$.  The
constants in \eqref{svw} have been chosen for later convenience.

In the type IIA limit where the distance between the tubes is much
larger than the size of the tubes, $\tau_2\gg 1$.  In this limit,
$q=e^{i\pi \tau}=e^{i\pi \tau_1-\pi \tau_2}$ is very small and, using
\eqref{qexpn}, we can approximate $F(z)$ as\footnote{Here, we have
dropped terms which are typically of order $q^2\sin(2\pi z)$.  The
modulus of this quantity is $|q^2\sin(2\pi z)| \sim e^{2\pi (|\Im
z|-\tau_2)}.$ As long as one stays in one fundamental region of the
$z$-torus, this is exponentially small for large $\tau_2$ and can be
safely dropped.}
\begin{align}
 F(z)&\approx \ln(e^{2\pi i z}-1).\label{F(z)approx}
\end{align}
Therefore, for example, we can approximate $s,v$ in \eqref{svw} as
\begin{equation}
\begin{split}
  s&\approx 2 \left[\log\left({e^{2\pi i (z-a_1)}-1 \over e^{2\pi i (z-a_2)}-1}\right)
 +i\pi(z-A-a)\right],\\
 v &\approx 2\pi i b_2\left[{1\over 1-e^{-2\pi i (z-a_1)}}-{1\over 1-e^{-2\pi i (z-a_2)}}
 +\half\right].
\end{split}\label{svapprox}
\end{equation}
We can similarly obtain an approximate form for $w$, but we do not
display it here explicitly because it is too lengthy.  On the other
hand, applying \eqref{qexpn} to \eqref{u2b2}, we see that, in the large
$\tau_2$ limit, $u_2\approx -\pi^2b_2^2.$ Classically, the two tubes
are sitting at the solution to $P_2(v)=0$, namely at $v=\pm
\sqrt{u_2}\approx \pm i\pi b_2$.  Therefore, if we want to keep the
distance between the two tubes to be finite, we must keep $b_2$
finite.

\begin{floatingfigure}[l]{0.35\textwidth}
\begin{center}
\epsfig{file=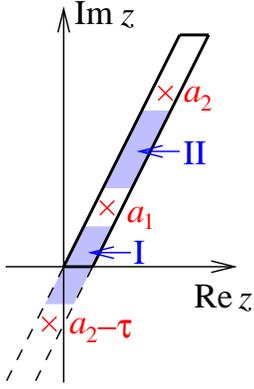,width=0.2\textwidth}
\caption{The $z$-torus for large $\tau_2$.  Region I ($\Im(z-a_1)\ll 0$,
$\Im(z-a_2)\ll 0$) and region II ($\Im(z-a_1)\gg 0$, $\Im(z-a_2)\ll 0$)
are depicted as shaded regions.}
\label{largetau2}
\end{center}
\end{floatingfigure}

In the type IIA limit, we take $R_{10}\to 0$ by definition.  What is the
relation between $\tau_2$ and $R_{10}$ in this limit?  For this, let us
look at the behavior of $s$ and $v$ near $z=a_{1,2}$, where they blow
up.  By examining \eqref{svapprox}, it is easy to see that, near
$z=a_{1,2}$,
\begin{align}
 s&\sim \pm 2\left[\log\left({2\pi i b_2\over v}\right)+\half i\pi a\right].
\end{align}
So, the distance between two NS5's at $v=v_0$ is given by
\begin{align}
 L&\equiv R_{10}\Re\bigl[s(z=a_1)-s(z=a_2)\bigr]
 =4R_{10}\left(\log\left|{2\pi b_2\over v_0}\right|-{\pi \tau_2\over 4}\right).
\end{align}
In other words,
\begin{align}
 |b_2|={|v_0|\over 2\pi}e^{{1\over 4}\left(\pi \tau_2-{L\over R_{10}M}\right)}.
\end{align}
In the IIA limit, we send $R_{10}\to 0$ keeping $L$ and $b_2$ finite.
Therefore, we must send $\tau_2\to\infty$ as $R_{10}\to 0$, as
\begin{align}
 \tau_2\sim {L\over \pi R_{10}}.\label{tau2scaleasgs}
\end{align}

\bigskip
If $\tau_2$ is very large, the $z$-torus becomes very long vertically
(along the $\Im z$ axis) as in Figure \ref{largetau2}.  We expect that
the long, narrow regions ``between'' marked points correspond to tubes
that descend to D4-branes in the IIA limit.
More specifically, let us call the region where $\Im(z-a_1)\ll 0$,
$\Im(z-a_2)\ll 0$ ``region I'' and the region where $\Im(z-a_1)\gg 0$,
$\Im(z-a_2)\ll 0$ ``region II'' (see Figure \ref{largetau2}).
Let us see that regions I and II correspond to two D4's in the IIA limit.

In region I, using \eqref{svapprox} and \eqref{F(z)approx}, one can
readily derive
\begin{equation}\begin{split}
 s&\approx 2\pi i (z+a-A),\\
 v&\approx i\pi b_2,\qquad\qquad\qquad\qquad\text{(region I)}\\
 w&\approx {4\pi i A_2\over \tau_2}(z+a-A).
\end{split}\end{equation}
So, if we move from $z=a_2-\tau$ (in the fundamental parallelogram below
our fundamental parallelogram) to $z=a_1$, then $s,v,w$ change
\emph{linearly} as
\begin{align}
\begin{array}{r@{\,}lcr@{\,}l}
 s&=-\half\pi i \tau &\to& s&=\half\pi i \tau, \\[.5ex]
 v&=i\pi b_2         &\to& v&=i\pi b_2,\\[.5ex]
 w&=-i\pi A_2        &\to& w&=i\pi A_2.
\end{array}
\end{align}
The change in $x^6$ is
\begin{align}
 \Delta x^6=R_{10}\Re(\Delta s)=-\pi R_{10} \tau_2=-L,
\end{align}
where in the last equality we used \eqref{tau2scaleasgs}.  Therefore,
region I indeed corresponds to a D4-brane at $v=i\pi b_2$ with length
$L$ along $x^6$.  Note also that this D4 is tilted along $w$ by $\Delta
w=2\pi i A_2$.
Similarly, one can show that region II corresponds to a D4-brane at
$v=-i\pi b_2$ with length $L$ along $x^6$, tilted along $w$ by $\Delta
w=-2\pi i A_2$.

Now let us turn to the regions near $z=a_{1,2}$, which must correspond
to NS5's curved along \eqref{superpotgt}.  Near $z=a_1$,
\begin{align}
 s&\approx 2\left[\log(2\pi i (z-a_1))+{i\pi \tau\over 4}\right],\qquad
 v\approx {b_2\over 2\pi i(z-a_1)}
 \qquad\qquad (z\approx a_1).
 \label{sv_near_z=a_1}
\end{align}
If we eliminate $z$,
\begin{align}
 v&=2\pi ib_2 e^{-{s\over 2}+{i\pi \tau\over 2}}.
\end{align}
Note that
\begin{align}
 \left|e^{-{s\over 2}+{i\pi \tau\over 2}}\right|
 =e^{-{x^6\over 2R_{10}}-{\pi \tau_2\over 4}}
 =e^{-{1\over 2R_{10}}(x^6+{L\over 2})}.
\end{align}
This goes to zero in the $R_{10}\to 0$ limit unless $x^6$ is very close to
$-L/2$, where the left NS5 sits.  If $x^6$ is very close to $-{L/2}$,
within $\CO(R_{10})$, then we can tune $x^6+{L/ 2}$ to allow $v$ to take
any value in $\bbC$.  Namely, we see that the region near $z=a_1$ indeed
corresponds to the left NS5 sitting at $x^6=-{L/ 2}$ and extending in
the $v$ direction.
How about the $w$ direction?  $w$ goes, near $z=a_1$, as
\begin{align}
 w&\approx {24A_6\over (z-a_{1})^5}+{2A_4\over (z-a_{1})^3}
 + {A_2\over z-a_{1}}
 \qquad\qquad (z\approx a_1).
 \label{w_near_z=a_1}
\end{align}
To see the relation between $v$ and $w$, we need the expression for $v$
up to lower powers in $(z-a_1)$ than was needed in
\eqref{sv_near_z=a_1}.  Near $z=a_1,$
\begin{align}
 v&\approx  2\pi i b_2\left[{1\over 1-e^{-2\pi i (z-a_{1})}}-\half\right]
 \qquad\qquad (z\approx a_1).
 \label{v_near_z=a_1}
\end{align}
If we eliminate $(z-a_1)$ from \eqref{w_near_z=a_1} and
\eqref{v_near_z=a_1}, we obtain
\begin{align}
 w&=
 {24A_6\over b^5}v^5+{2(A_4+20\pi^2 A_6)\over b_2^3}v^3
 +{A_2+2\pi^2A_4+16\pi^4A_6 \over b_2}v+\CO\left({1\over v}\right).
\end{align}
One can readily see that this is the expected behavior
\eqref{superpotgt}, if one uses \eqref{Asbdrycond} and that
$\wp(\tau/2)\approx -\pi^2/3$, $g_2\approx 4\pi^4/3$ for large $\tau_2$.

Similarly, one can show that the region near $z=a_2$ corresponds to an
NS5 at $x^6=L/2$ along $w=-W'(v)$.

So far we have been treating the parameters $A_{2,4,6},b_2,\tau_2$ as if
they were all arbitrary, but they are actually subject to the constraint
\eqref{compA2eqn}.  For large $\tau_2$, this equation reduces to
\eqref{tau2scM}:
\begin{align}
 \tau_2&={2\over \pi}{W'(v)\over v W''(v)}.
 \label{muxb5Dec07}
\end{align}
where $v=\pm i\pi b_2$.  An important difference from the semiclassical
limit considered in section \ref{sec:mechanism} is that we send
$\tau_2\to\infty$ as \eqref{tau2scaleasgs} as we take $R_{10}\to 0$.  This
means that
\begin{align}
 {v W''(v)\over W'(v)}={2\over\pi \tau_2}={2R_{10}\over L}\to 0.
\end{align}
Therefore, in the IIA limit, the D4's must sit at a critical point of
$w(v)=W'(v)$, {\it i.e.}, it must be that $w'(v)=0$.  This is because
in the IIA limit the log bending of NS5's along $x^6$ disappears and
hence there is no force to cancel the $w$-tilting force.  So, the D4's
must sit at points where they can sit in equilibrium without any force
(although unstable in this limit), which happens for $w'(v)=0$. 


\end{document}